\documentclass[prl,aps,10 pt,two column]{revtex4-1}
\usepackage{graphicx}% Include figure files
\usepackage{multirow}
\usepackage{threeparttable}
\usepackage{dcolumn}% Align table columns on decimal point
\usepackage{bm}% bold math
\usepackage{mathtools,amssymb}
\usepackage{upgreek}
\usepackage{color}
\usepackage{float}
\makeatletter

\newcommand{\Rmnum}[1]{\expandafter\@slowromancap\romannumeral #1@}

\makeatother

\begin{document}
\title {Spontaneous time reversal symmetry breaking at individual grain boundaries in graphene}

\author{Kimberly Hsieh$^{1,*}$, Vidya Kochat$^{1,5,*}$,  Tathagata Biswas$^{1,6}$, Chandra Sekhar Tiwary$^{2,7}$, Abhishek Mishra$^3$, Gopalakrishnan Ramalingam$^4$, Aditya Jayaraman$^1$, Kamanio Chattopadhyay$^2$, Srinivasan Raghavan$^{3,4}$, Manish Jain$^1$ \& Arindam Ghosh$^{1,3}$}
\vspace{1.5cm}
\affiliation{$^1$ Department of Physics, Indian Institute of Science, Bangalore 560 012, India}
\affiliation{$^2$ Department of Materials Engineering, Indian Institute of Science, Bangalore 560 012, India}
\affiliation{$^3$ Centre for Nano Science and Engineering, Indian Institute of Science, Bangalore 560 012, India}
\affiliation{$^4$ Materials Research Center, Indian Institute of Science, Bangalore 560 012, India}

\thanks{These authors contributed equally to this work.}

\altaffiliation{\\$^5$ Present address: Materials Science Centre, Indian Institute of Technology Kharagpur, Kharagpur 721302, India}
\altaffiliation{\\$^6$ Present address: Department of Physics, Arizona State University, Tempe, AZ 85287, USA}
\altaffiliation{\\$^7$ Present address: Department of Metallurgical and Material Engineering, Indian Institute of Technology Kharagpur, Kharagpur 721302, India\\}

\begin{abstract}
                               
%Nearly dispersionless flat electronic bands in graphene-based system is known to host a range of exciting properties. Apart from the zigzag edge and the critically mis-aligned twisted bilayers of graphene, such bands can also appear in the grain boundaries (GBs) of polycrystalline graphene, but their experimental signature remains unexplored. 
Graphene grain boundaries have attracted interest for their ability to host nearly dispersionless electronic bands and magnetic instabilities. Here, we employ quantum transport and universal conductance fluctuations (UCF) measurements to experimentally demonstrate a spontaneous breaking of time reversal symmetry (TRS) across individual GBs of chemical vapour deposited graphene. While quantum transport across the GBs indicate spin-scattering-induced dephasing, and hence formation of local magnetic moments, below $T\lesssim 4$~K, we observe complete lifting of TRS at high carrier densities ($n \gtrsim 5\times 10^{12}$cm$^{-2}$) and low temperature ($T\lesssim 2$~K). An unprecedented thirty times reduction in the UCF magnitude with increasing doping density further supports the possibility of an emergent frozen magnetic state at the GBs. Our experimental results suggest that realistic GBs of graphene can be a promising resource for new electronic phases and spin-based applications. 
               
%Flat band-induced magnetism at graphene grain boundaries (GBs)  has been of great theoretical interest for over a decade, but its experimental signatures remain largely unexplored. Restricted spatial dimension of the GBs, and weak inter-spin exchange energy have made conventional magnetic probing a challenging task. Here, we employ quantum transport and universal conductance fluctuations (UCF) measurements to experimentally demonstrate the spontaneous breaking of time reversal symmetry (TRS) across individual GBs of chemical vapour deposited graphene. At high carrier densities ($n \gtrsim 5\times 10^{12}$cm$^{-2}$), quantum transport suggests spin scattering-induced dephasing at temperatures $T\lesssim 4$~K, while UCF measurements reveal complete lifting of TRS at $T\lesssim 2$~K. Additionally, an unprecedented thirty times reduction in the UCF magnitude at the GBs is observed as a function of carrier density, which can be explained through local moments interacting via the Ruderman-Kittel-Kasuya-Yosida (RKKY) mechanism. The emergence of electrically tunable magnetism in graphene GB systems makes them useful candidates for spintronic applications employing novel 2D magnetic circuits.

\end{abstract}  

\maketitle

\begin{figure}[!b]
\begin{center}
\includegraphics [width=1\linewidth]{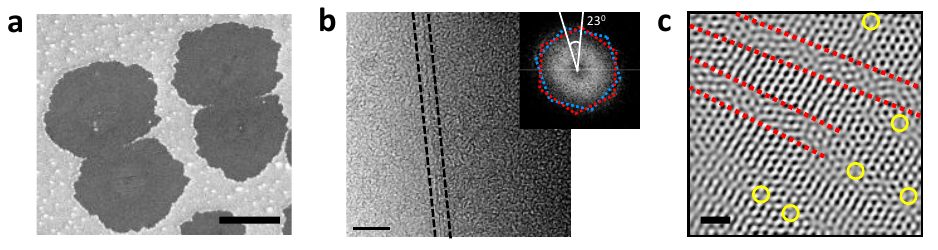}
\end{center}
\vspace{-0.4cm}
\caption{\textbf{a},~Scanning electron micrograph of a typical pair of graphene grains with grain size$\approx 15$~$\upmu$m forming a GB in between. Scale bar, 10~$\upmu$m \textbf{b},~Bright field TEM image of the GB formed between two grains. Selected area electron diffraction (SAED) pattern in inset shows the misorientation angle between the grains $\approx23^{\circ}$. Scale bar, 5~nm. \textbf{c}~HRTEM image of the GB region where line and point defects are outlined. Scale bar, 1~nm.}
\label{F1}
\vspace{-0.4cm}
\end{figure}

%\textbf{a},~Schematic showing two categories of defects in graphene: point and extended defects. \textbf{b,c},~Bandstructure and density of states calculations for graphene with GB containing \textbf{b},~pentagon-octagon defects (GB$(2,0)|(2,0)$) and \textbf{c},~pentagon-heptagon defects (GB$(5,0)|(3,3)$) showing enhanced DOS due to flat band formation. 

Structural disorder in graphene originates from defects classifiable into two categories - point defects (vacancies, Stone-Wales defects) and extended defects such as grain boundaries (GBs). Vacancies result in localized states close to zero energy leading to magnetic moment formation in graphene, experimentally confirmed by the observation of spin-split resonances in scanning tunneling microscopy (STM) at monovacancies~\cite{Ugeda_PRL2010}, measurements of spin currents~\cite{McCreary_PRL2012} and possibility of Kondo effect~\cite{Chen_NatPhys2011,Jiang_NatCommun2018}. GBs lead to local modification of graphene band structure by introducing weakly dispersing, nearly flat electronic bands with enhanced density of states (DOS), either at zero energy (translational GB$(2,0)|(2,0)$)~\cite{Kou_ACSNano2011,Alexandre_NanoLett2012} or finite energies (tilt GB$(5,0)|(3,3)$)~\cite{Dutta_SciRep2015,Nemes_Carbon2013,Luican_2DMater2016}. While preliminary studies projected GBs as detrimental to electronic transport~\cite{Yu_NatMat2011,Gargiulo_NanoLett2013,Yazyev_NatMater2010,Tuan_NanoLett2013,Koepke_ACSNano2013,Clark_ACSNano2013,Tapaszto_APL2012,Yu_NatMat2011,Jauregui_SSC2011}, successive experiments showed that these drawbacks can be overcome by tailoring the growth conditions~\cite{Tsen_Science2012,Vidya_Nanoletters2016}. Magnetotransport measurements across isolated GBs reveal enhanced weak localization (WL) compared to single-crystalline grains, indicating stronger intervalley carrier scattering due to lattice disorder~\cite{Yu_NatMat2011,Jauregui_SSC2011}. However, a comprehensive study of the symmetry-breaking mechanisms at graphene GBs through direct measurements of universal conductance fluctuations (UCF) in the inter- and intra-grain regions has so far been lacking. 

Magnetic ordering has been predicted at GBs, either by localization at non-trivially coordinated C-rings~\cite{Dutta_SciRep2015} or assisted by strain, {\it e.g.} in translational line defects with octagon-pentagon pairs~\cite{Kou_ACSNano2011,Alexandre_NanoLett2012}. In realistic GBs realized during chemical vapour deposition~(CVD) growth, nucleation centers grow independently and fuse in local bonding environment. Such GBs comprise multiple defect realizations, including vacancies, Stone-Wales defects, intermittent 1D line defects~\cite{Yazyev_NatMater2010,Lahiri_NatNano2010}, multi-membered C-rings~\cite{Huang_Nature2011,Dutta_SciRep2015} etc., causing strong increase in charge carrier scattering and electronic noise~\cite{ Tsen_Science2012,Koepke_ACSNano2013,Vidya_Nanoletters2016}. Despite both numerical~\cite{Kou_ACSNano2011,Alexandre_NanoLett2012,Dutta_SciRep2015} and spectroscopic~\cite{Cervenka_NatPhys2009,Nemes_Carbon2013,Luican_2DMater2016,Lahiri_NatNano2010} evidences of large enhancement in local DOS and spin-splitting, no tangible impact of {\it e-e} interaction at the GBs has so far been observed. This work combines quantum transport and UCF to probe local charge and spin excitations across individual graphene GBs. The UCF magnitude, determined by the symmetry of the underlying Hamiltonian via the Wigner-Dyson parameter $\beta$~\cite{Beenakker_RevModPhys1997}, reveals a full spontaneous lifting of the time reversal symmetry (TRS) in the GB region for $T\lesssim 2$~K and $n\gtrsim 5 \times 10^{12}$~cm$^{-2}$. The temperature and density-dependence of UCF links the TRS lifting to a frozen magnetic state arising from the GB defect sites.

\begin{figure}[!t]
\begin{center}
\includegraphics [width=1\linewidth]{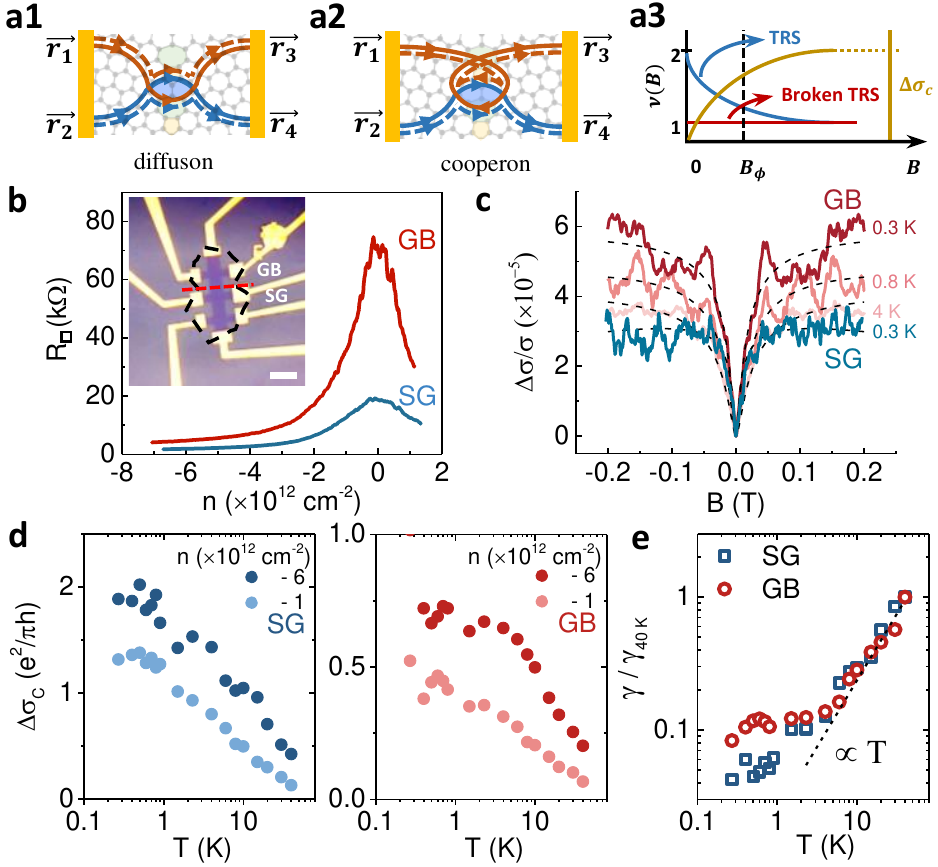}
\end{center}
\vspace{-0.4cm}
\caption{\textbf{a},~Schematic showing \textbf{a1},~a pair of crossings representing diffusons,  \textbf{a2}, a pair of crossings representing Cooperons, and \textbf{a3},~the expected behaviour of $\nu(B)$ as a function of $B$ for TRS-invariant and TRS-broken systems. \textbf{b},~Sheet resistance~($R_\Box$) as a function of gate voltage~($V_{\mathrm{BG}}$) for intra-grain~(SG) and inter-grain~(GB) regions of D1 at $T=0.3$~K. (Inset) Optical micrograph of a typical device. The morphology of the original pair of coalesced grains is shown (black line) along with the approximate GB location (red line). Scale bar, 10~$\mu$m.  \textbf{c},~Magnetoconductance measurements are shown for $n=-6\times10^{12}$~cm$^{-2}$ at $T=0.3$~K for the SG region and $T=0.3$,~$0.8$, and $4$~K for the GB region, clearly exhibiting WL.  Dashed lines correspond to HLN fits. \textbf{d},~Quantum correction to conductivity $\Delta\sigma_{c}$ in units of $e^2/\pi h$ plotted for both SG and GB regions as a function of $T$ for $n=-1\times 10^{12}$~cm$^{-2}$ and $-6\times 10^{12}$~cm$^{-2}$. \textbf{e},~Scattering rate $\gamma$ normalized to its value $\gamma_{40\,\mathrm{K}}$ at $T=40$~K,  plotted versus $T$ for the SG and GB regions at $n=-6\times10^{12}$~cm$^{-2}$ is shown, where the black dotted line indicates the temperature regime where Nyquist scattering dominates.}
\label{F2}
\vspace{-0.4cm}
\end{figure}

%which we find to be robust against accompanying random disorder (DFT calculations for a realistic, disordered GB of finite width~\cite{Koepke_ACSNano2013,Clark_ACSNano2013} shown in Fig.~S10 of Supplementary Information (SI)),

%\section{Results and Discussion}

We measured three devices (D1, D2 and D3) from CVD-synthesized graphene (Supplemental Material (SM) section S1), optimized to ensure partial fusion of crystallites (scanning electron micrograph in Fig.~\ref{F1}a). High-resolution transmission electron microscopy (HRTEM) (Fig.~\ref{F1}b) performed on a pair of similarly synthesized grains reveal an average width $\sim 10$~nm of the disordered region, and a misorientation angle $\approx 23^{\circ}$ between the parent crystallites. The GBs form a highly disordered region consisting of arrays of line dislocations and under-coordinated C-atoms (Fig.~\ref{F1}c) similar to that observed in STM and TEM studies~\cite{Cervenka_NatPhys2009,Nemes_Carbon2013,Luican_2DMater2016,Lahiri_NatNano2010,Vidya_Nanoletters2016}.

Fig.~2a schematically describes the conceptual basis of our experimental approach. The quantum interference effect, that underpins both quantum correction to conductivity ($\Delta\sigma_c$) and UCF ($\langle\delta G_{\phi}^2\rangle$), depends on crossings of time-reversed path pairs as the electron (or hole) diffuses across the sample over $\tau_D = L^2/D$, the Thouless time, where $L$ and $D$ are the length of the system and carrier diffusivity, respectively. While $\Delta\sigma_c$ is determined by the probability of single self-crossing, the correlation function in $\langle\delta G_{\phi}^2\rangle \sim \langle G(0)G(\tau)\rangle_\tau$ requires two spatially separated crossing points (thus involving larger number of defect sites), thereby defining closed loops encircled either in the same (diffusons) or opposite (Cooperons) senses with identical structure factors (Figs. 2a1 and 2a2). This has two important consequences: first, compared to $\Delta\sigma_c$ ($\sim \ln(\tau_D\gamma)$), the UCF magnitude $\langle\delta G_{\phi}^2\rangle \sim (\tau_D\gamma)^{-2}$ is exponentially more sensitive to emergent dephasing processes in two dimensions, where $\gamma$ is the dephasing rate, and thus a more suitable tool when the dephasing processes are confined within spatially restricted regions such as the GBs. Second, when TRS is lifted, usually by a transverse magnetic field $B \gg B_\phi$, $B_\phi$ being the field corresponding to one flux quantum threading a phase coherent cell, the Cooperon contribution is removed, decreasing $\langle\delta G_{\phi}^2\rangle$ {\it exactly} by a factor of two. The reduction factor is protected by the symmetry of the underlying Hamiltonian, {\it i.e.} $\langle\delta G_{\phi}^2\rangle \sim (e^2/h)^2/\beta$, where $\beta = 1$ for time reversal invariant systems (orthogonal ensemble), and $\beta = 2$ when TRS is absent (unitary ensemble). For a time-reversal invariant system~\cite{Stone_PRB1989}, a crossover function $\nu(B)$ defined as

\begin{equation}\label{Crossover}
\nu(B)=\frac{N(B)}{N_{\phi}}=1+\frac{2}{b^{2}}\sum_{n=0}^{\infty}\frac{1}{[\left(n+\frac{1}{2}\right)+\frac{1}{b}]^{3}} 
\end{equation}

\noindent where $b=8\pi B(l_{\phi})^{2}/(h/e)$ is the dimensionless magnetic field which captures the reduction in UCF as a function of $B$ (Fig.~2a3). Here, $N(B)$ and $N_\phi$ are the values of $\langle\delta G_{\phi}^2\rangle$ at $B$ and at $B \gg B_\phi$, respectively. When TRS is spontaneously removed (magnetic systems), $\nu(B)$ remains unaffected at the scale of $B_\phi$, as observed in ferromagnetic films~\cite{Lee_PRB2004}.

For electrical transport, the graphene grains were transferred on to Si/SiO$_2$ substrates, patterned into Hall bars such that measurements across the GB and within single grain (SG) can be carried out simultaneously (Fig.~2b inset). %Due to strong intrinsic electron doping ($V_{\mathrm{CNP}} \approx 40$~V for D1), we achieved doping $n$ as large as $-7\times10^{12}$~cm$^{-2}$ in the hole side in D1. 
The excess disorder in the GB region results in enhancement in the resistivity by a factor of $\sim 2 - 5$ times that of the SG region depending on $n$ (Fig.~2b) and a consequent suppression of the carrier mobility ($\mu_{\mathrm{SG}}\approx 480$~cm$^2$V$^{-1}$s$^{-1}$ while $\mu_{\mathrm{GB}}\approx 220$~cm$^2$V$^{-1}$s$^{-1}$). Magnetotransport measurements down to $T = 0.3$~K for a fixed $n$ ($\approx -6\times10^{12}$~cm$^{-2}$) indicate enhanced WL correction at the GB region (Fig.~2c), signifying stronger intervalley scattering from short range lattice defects~\cite{Yu_NatMat2011,Jauregui_SSC2011,Pal_PRL2012}. Fitting (dashed lines in Fig.~2c) the modified Hikami-Larkin-Nagaoka (HLN) expression for graphene~\cite{McCann_PRL2006} to magnetoconductance yields both the quantum correction to conductivity $\Delta \sigma_c$ and the dephasing length $l_{\phi}$ (Fig.~S2a). The $T$-dependence of $\Delta\sigma_c$ in Fig.~2d shows that the quantum correction behaves differently between SG (left) and GB (right) regions, especially at high $n$. In both cases, we find $\Delta \sigma_c \propto \ln(T)$ at low $n$, as expected for diffusive non-magnetic conductors where dephasing takes place via Nyquist scattering from {\it e-e} interaction so that $\gamma =D/l_{\phi}^2\propto T$. Direct evaluation of $\gamma$ (Figs.~2e and S2b) from HLN fits confirm this proportionality with $T$. The quantum correction in the GB region at high $n$~($\gtrsim 5\times 10^{12}$~cm$^{-2}$) however deviates from this behaviour, where we find both $\Delta \sigma_c$ (Fig.~2d, right panel) and $\gamma$ (Fig.~2e) saturate below $T\sim 7$~K. 
%We eliminate effects of elevated electron temperature as probable cause for the saturation in these parameters in the GB region, because both SG and GB regions were integrated in the same device and the excitation bias $V_{\mathrm{SD}} \lesssim k_{\mathrm{B}}T/e$). 

Low temperature saturation of $\gamma$ in metals is often attributed to inelastic processes from spin-flip scattering~\cite{Pierre_2001,Pierre_PRL2002,Pierre_PRB2003}. Neglecting electron-phonon scattering at such temperatures~\cite{Morozov_PRL2008,Tikhonenko_PRL2009}, we can write $\gamma=\gamma_{ee}+\gamma_s$, where $\gamma_{ee}$ is the {\it e-e} induced scattering rate and $\gamma_s$ is the spin-flip scattering rate due to dilute magnetic impurities. The observed saturation in $\gamma$ at temperatures $T\lesssim 7$~K can thus be explained from increasing Nyquist scattering ($\gamma_{ee}\propto T$) countering the reduction in $\gamma_s$ above the Kondo temperature $T_{\mathrm{K}}$~\cite{Maple_1973,Haesendonck_PRL1987}. Such an anomalous $T$-dependence of $\gamma$ in the inter-grain region hints at the formation of magnetic moments that can interact at lower temperatures leading to frozen magnetic ordering~\cite{Schopfer_PRL2003}. However, the competing effects of localization and anti-localization due to graphene's chiral charge carriers makes it ambiguous to detect or claim such possibilities using WL alone.

\begin{figure}[!t]
\begin{center}
\includegraphics [width=1\linewidth]{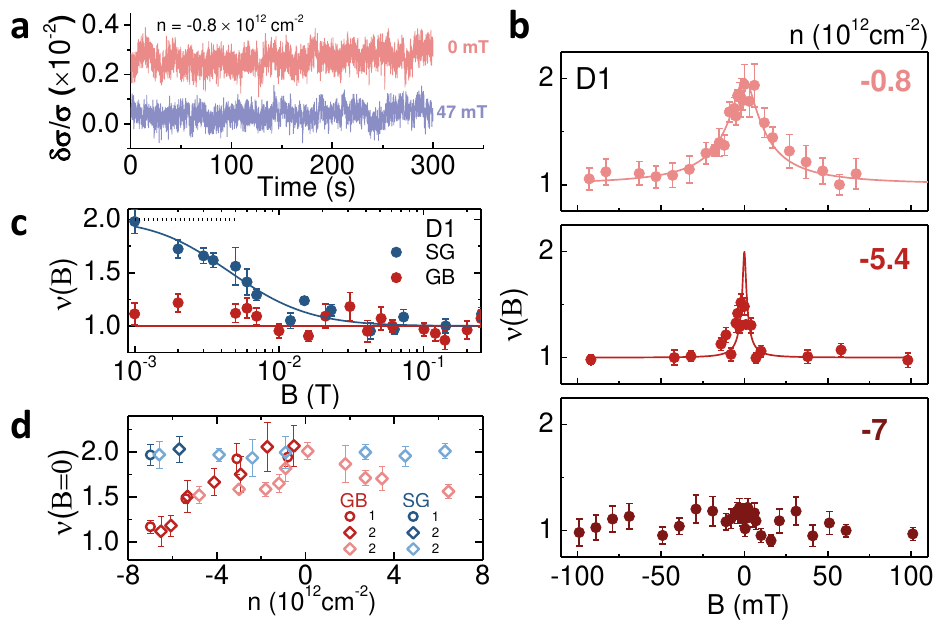}
\end{center}
\vspace{-0.4cm}
\caption{\textbf{a}, Conductivity fluctuations for GB region of D1 at $n=-0.8\times10^{12}$~cm$^{-2}$ for $B=0$~mT (pink) and $B=47$~mT (purple), clearly indicating a reduction in the fluctuation magnitude at $B\gg B_{\phi}$. \textbf{b},~$\nu(B)$ plotted for three different $n$ at $T=0.3$~K for D1 showing spontaneous TRS breaking at zero field as $n$ is increased. Solid lines are fits to Eq.~\ref{Crossover}. \textbf{c},~$\nu(B)$ for the SG and GB regions plotted at $T=0.3$~K for D1 at $n=-7\times10^{12}$~cm$^{-2}$ indicating that spontaneous TRS breaking occurs only in the presence of a GB. \textbf{d},~Noise reduction factor $\nu(B=0)$ for the SG and GB regions of D1 (circles) measured at $T=0.3$~K (darker) and D2 (diamonds) at $T=0.3$~K (darker) and $T=4.5$~K (lighter) as a function of $n$.}
\label{F3}
\vspace{-0.4cm}
\end{figure}

To complement quantum transport, we carried out UCF measurements in two different ways: (1) From slow time-dependent fluctuations in the conductance relating directly to the ensemble fluctuations of disorder configuration via ergodic hypothesis~\cite{Stone_PRB1989, Birge_PRB1990,Feng_PRL1986,Saquib_PRL2014} (Fig.~3, SM section S4), and (2) by analyzing the reproducible and aperiodic fluctuations in $G$ by tuning the Fermi energy (Fig.~4b, SM section S5). %In the first method, the evaluation of the variance $\langle\delta G_{\phi}^2\rangle$ was obtained from statistically meaningful windows of time ($\approx 1200$~seconds with $\approx 18750$ realizations of $G$) while the second method involved overlapping $4$~V intervals in $V_{\mathrm{BG}}$, each with $800$ realizations of $G$. The equivalence of the two methods has been detailed elsewhere~\cite{Stone_PRB1989,Pal_PRL2012}.
The time-dependent conductance fluctuations across the GB of D1 at $n=-0.8\times{10}^{12}$~cm$^{-2}$ is plotted in Fig.~3a, clearly displaying a reduction in the relative magnitude of fluctuations at $B=0$~T and $B = 47$~mT~($\gg B_{\phi}$). Fig.~3b shows the $B$-dependence of $\nu(B)$, defined in Eq.~\ref{Crossover}, from $\langle\delta G_{\phi}^2\rangle$ evaluated from time-dependent conductance fluctuations in device D1 for three different $n$ at the GB region. At low $n$ ($\approx 0.8\times10^{12}$~cm$^{-2}$), $\nu(B)$ shows a clear factor-of-two reduction as $B$ increases beyond $\sim 30$~mT, which corresponds to $B_\phi$ (Fig.~3b, uppermost panel). This suggests TRS to be preserved in GB regions at low $n$, similar to 2D systems such as exfoliated graphene~\cite{Pal_PRL2012}, topological insulators~\cite{Islam_PRB2018}, doped Si/Ge systems~\cite{Saquib_SciRep2017} and non-magnetic films~\cite{Birge_PRB1990}. %lithium wires~\cite{Moon_PRB1996}, GaAs/AlGaAs heterostructures~\cite{Debray_PRL1989,Mailly_EurPhys1989},  silver thin films~\cite{McConville_PRB1993}, bulk doped Si~\cite{Ghosh_PRL2000} and AuPd wires~\cite{Trionfi_PRB2004,Trionfi_PRB2005}. 
However, with increasing $n$, a progressive reduction in $\nu(B)$ at $B=0$ was observed across the GB approaching unity, and thus $B$-independent $\nu$, for $n\gtrsim-6\times{10}^{12}$~cm$^{-2}$ (Fig.~3b, bottom panel). The insensitivity of $\nu(B)$ to transverse field at the scale $B\sim B_{\phi}$ is a unique characteristic of systems with spontaneously broken TRS, as observed before in ferromagnetic films~\cite{Lee_PRB2004} and lightly-doped semiconductors in strongly interacting regime~\cite{Saquib_PRL2014}. A similar trend was observed for D2 (Figs.~3d, S5 and S6) where the reduction in $\nu(B)$ was observed in both doping regimes. Remarkably, the spontaneous breaking of TRS was observed only in the inter-grain region, while the intra-grain region continues to show a factor-of-two reduction in UCF magnitude with $B$ at similar high densities (Fig.~3c, higher $T$ in Fig.~S9). The near $B$-independence of $\nu$ at high $n$ was found to be ubiquitous to quantum transport across GBs in CVD graphene as shown for D1 in Fig.~3b and D2 in Fig.~S5. The solid lines in Figs.~3b and c correspond to fits of $\nu(B)$ according to Eq.~\ref{Crossover} with $l_{\phi}$ as the only fitting parameter. %The values of $l_{\phi}$ obtained from WL and UCF measurements agree within a factor of two for all devices. 
Fig.~3d shows the noise reduction factor $\nu(B=0)=N(B=0)/N_{\phi}$ for D1 and D2 measured at $T=0.3$ and $4.5$~K as a function of $n$. At $T=4.5$~K, the minimum $\nu(B)$ ($\approx 1.5$) at highest experimental $n$ indicate only partial removal of TRS. The factor of two reduction of $\nu(B)$ across the SG region was maintained throughout the entire density range, implying that TRS is lifted solely in the presence of the GB. 

\begin{figure}[!t]
\begin{center}
\includegraphics [width=1\linewidth]{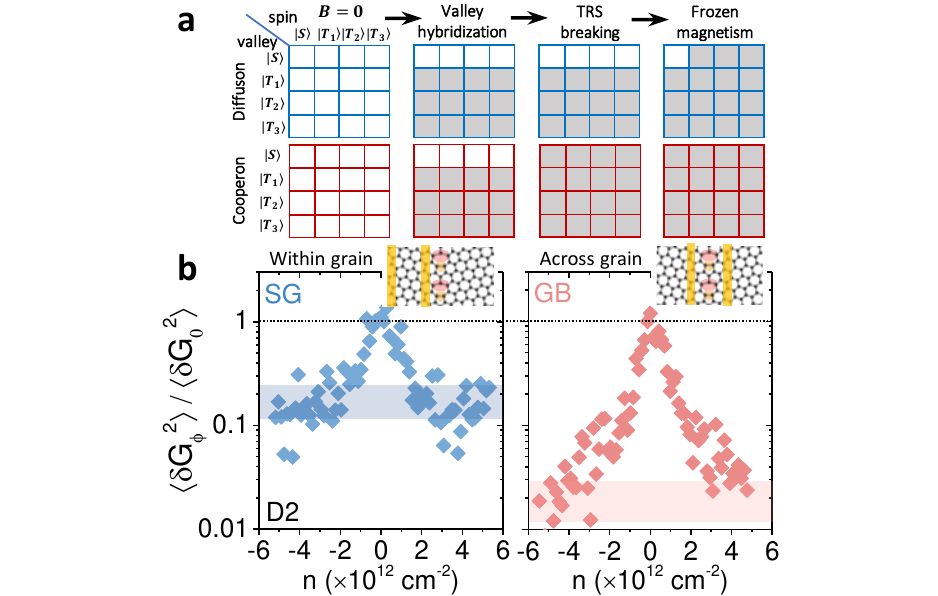}
\end{center}
\vspace{-0.4cm}
\caption{\textbf{a},~Schematic describing the contribution of diffuson (blue) and cooperon (red) singlet ($|S\rangle$) and triplet ($|T_1\rangle$, $|T_2\rangle$ and $|T_3\rangle$) states to the UCF magnitude $\langle \delta G_{\phi}^2 \rangle$ in different symmetry classes. Valley hybridization leads to a factor of four reduction in $\langle \delta G_{\phi}^2 \rangle$ while magnetic impurities reduces $\langle \delta G_{\phi}^2 \rangle$ by a {\it further} factor of eight due to gapping of spin diffuson triplets and all cooperons. \textbf{b},~The variance in conductance $\langle \delta G_{\phi}^2 \rangle$ within a phase-coherent box of $l_{\phi}^2$ normalized to its value at the Dirac point $\delta\langle G_0^2 \rangle$ as a function of $n$ at $T=0.3$~K, showing a factor of $\approx 4$ reduction in the SG region and a factor of $\approx 30$ reduction in the GB region.}
\label{F4}
\vspace{-0.4cm}
\end{figure}

To understand the origin of TRS breaking in the GB region, we then measured the $n$-dependence of the zero-$B$ magnitude of UCF, which can distinguish between TRS breaking from external $B$ field and that from an emergent frozen magnetic state~\cite{Akkermans_2007}. For this, $\langle\delta G_{\phi}^2\rangle$ was calculated from reproducible fluctuations in $G$ within small windows of $V_{\mathrm{BG}}$ i.e. from $E_{\mathrm{F}}$ (SM section S5). The SG region exhibits a factor of $\approx 4$ reduction in $\langle\delta G_{\phi}^2\rangle$ (Fig.~4b, left panel) due to valley symmetry lifting thereby suppressing the UCF from valley triplet channels, a behaviour observed in exfoliated graphene~\cite{Pal_PRL2012}.
%Tuning $n$ away from $V_{\mathrm{CNP}}$ results in a crossover from long-range Coulomb scattering to short-range scattering resulting in valley hybridization and gapping of the pseudospin triplet states of both diffusons and cooperons (see schematic in Fig.~4a)~\cite{Pal_PRL2012}. Consequently, a factor of $\approx 4$ reduction in the UCF magnitude  $\langle\delta G_{\phi}^2\rangle$ normalized to its value $\langle\Delta G_0^2\rangle$ at $V_{\mathrm{CNP}}$ is observed for the intra-grain region (left panel of Fig.~4b). 
In contrast, the UCF magnitude in the GB region exhibits a drastic reduction (Fig.~4b, right panel) by a factor of $\approx 30$ as $n$ is increased. 
%which may be qualitatively explained by the suppression of conductance fluctuations through spin-flip processes effectively changing the disorder realization, leading to self-averaging. Quantitatively, the suppression in $\langle\delta G_{\phi}^2\rangle$ as a function of $n$ can be explained from the combined effects of intervalley scattering (factor of four) and scattering at {\it frozen} magnetic impurities (factor of eight) at the GBs (see schematic in Fig.~4a).
This unique and unprecedented reduction can be quantitatively understood from a combination of valley hybridization, TRS breaking and suppression of spin triplet channels in the presence of static (measurement time short compared to Korringa relaxation time) spin-dependent scattering, as depicted schematically in Fig.~4a. The static spin texture, or `frozen magnetic state', at large $n$ may happen when the defect-bound magnetic impurities interact via RKKY (Ruderman-Kittel-Kasuya-Yosida) exchange, forming long or short ({\it e.g.} a spin glass) range spin-ordered states~\cite{Anderson1978,Schopfer_PRL2003}. Thus, the UCF measurement in graphene containing a GB suggests a rather unexpected effect of doping, which manifests in both valley and (static) spin polarization when carrier density is made sufficiently large.

\begin{figure}[!t]
\begin{center}
\includegraphics [width=1\linewidth]{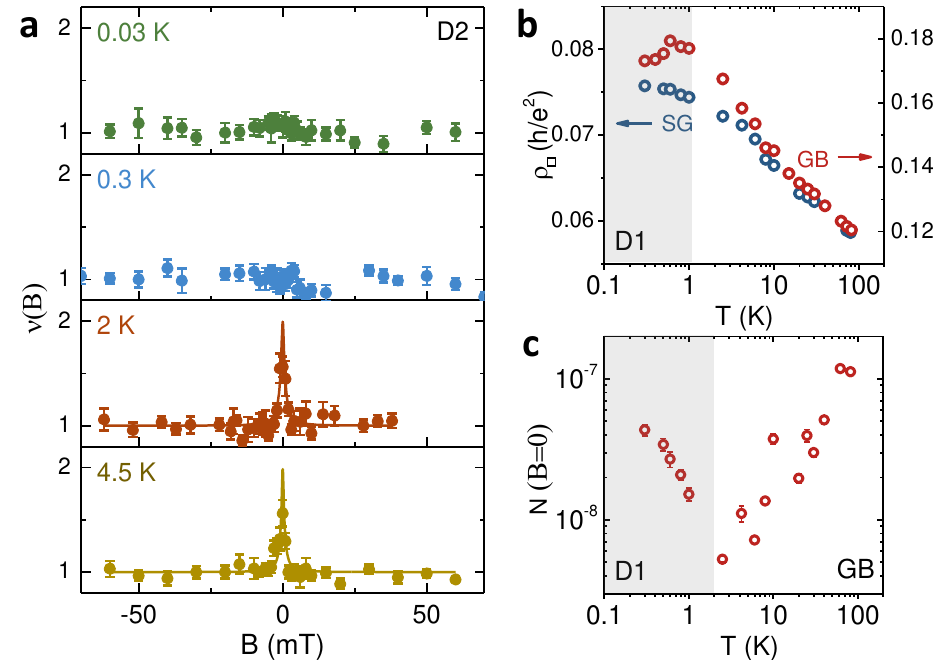}
\end{center}
\vspace{-0.4cm}
\caption{\textbf{a},~$\nu(B)$ plotted for D2 at $n=6.1\times10^{12}$~cm$^{-2}$ showing spontaneous TRS breaking at zero field only at temperatures $T\lesssim 2$~K. \textbf{b},~$T$-dependence of the reduced sheet resistance $\rho_{\square}$ (in units of $h/e^2$) averaged from resistance fluctuations measurements at $B=0$~T for SG (blue) and GB (red) regions at $n=-6\times10^{12}$~cm$^{-2}$ for D1. \textbf{c},~Normalized variance $N=S_\sigma/\sigma^2$ at $B=0$ as a function of temperature is plotted for GB region of D1 at $n=-6.9\times10^{12}$~cm$^{-2}$ clearly indicating a sharp increase in $N(B=0)$ at lower temperatures.}

%\textbf{b},~Experimental demonstration of defect-induced distortion on the parallel lattice array of a representative HRTEM image of a GB. \textbf{c},~Histogram obtained from the HRTEM image in \textbf{b} revealing an average defect distance $R\approx 2$~nm. \textbf{d},~The reduction factor $\nu(B=0)$ of the time-dependent UCF in the intergrain region plotted as a function of temperature $T$ at high electron density $n=6.1\times10^{12}$~cm$^{-2}$.
\label{F5}
\vspace{-0.4cm}
\end{figure}

%The saturation in low-temperature scattering rates and the direct observation of spontaneous TRS breaking at the GB upon doping, signals the possible development of short/long range magnetic order at the graphene GBs, likely through the formation and subsequent interaction of local moments at GB defect sites. %The Fermi level becomes pinned at these GB localized states and upon substantial doping, increased electron-electron interactions in the narrow band leads to a Stoner instability. The spin flip scattering time $\tau_s\propto\tau_p$, the momentum relaxation time, suggesting Elliot-Yafet spin scattering mechanism at the GB, which is the dominant spin relaxation mechanism in monolayer graphene~\cite{Elliott_PR1954}. 
To estimate the energy scale for local moment interaction at the GB, we study the effect of temperature on $\nu(B)$. The normalized magnetonoise $\nu(B)$ for the GB region in device D2 at $n=6.1\times{10}^{12}$~cm$^{-2}$ with varying temperature is shown in Fig.~\ref{F5}a (data at higher $T$ in Fig.~S7a). Evidently, the spontaneous TRS breaking occurs only at temperatures $\lesssim 2$~K, while $\nu(B)$ approaches $\sim 2$ as $T$ is increased (data for D2 at $T=4.5$~K in Fig.~S6 and D3 at $T=8$~K in Fig.~S8). To estimate the exchange interaction between moments, we first estimate the Kondo temperature $T_{\mathrm{K}}\simeq 20$~K from the $T$-dependence of the sheet resistance $R_{\square}$ at finite $B$ where WL corrections are suppressed (Fig.~S14). This $T_{\mathrm{K}}$ is compatible with studies on irradiated graphene~\cite{Chen_NatPhys2011,Jiang_NatCommun2018}. The RKKY interaction between moments can be estimated as~\cite{Sherafati_PRB2011_RKKY} $k_{\mathrm{B}} T_{\mathrm{RKKY}}\sim 27 j^2a^4/64 \pi v_{\mathrm{F}}\hbar R^3\sim 2.2$~K (lattice constant $a\approx 0.246$~nm and Fermi velocity in graphene $v_{\mathrm{F}}\approx10^6$~m~s$^{-1}$), where the Kondo exchange $j\approx 2.3$~eV was obtained from the experimental $T_{\mathrm{K}}$ and the DOS in the GB region, $D(E_{\mathrm{F}})\sim 0.05$~eV$^{-1}$ (Fig.~S12)~\cite{Kou_ACSNano2011,Alexandre_NanoLett2012,Dutta_SciRep2015}. Such a large $j$ value agrees with previous theoretical calculations~\cite{Sengupta_PRB2008,Cazalilla_Arxiv2012,Mitchell_PRB2013}.
%and can be attributed to GBs being fully immersed within the graphene lattice, contributing directly to the local DOS unlike transition metal adatoms which tend to reside above the graphene plane~\cite{Wang_NanoLett2012}. 
The average defect distance $R\approx 2$~nm can be estimated from the HRTEM image of the GB region on alignment in two-beam condition (SM Fig.~S13a with corresponding histogram in Fig.S13b). This value of $T_{\mathrm{RKKY}}$ agrees reasonably well with the $T$ dependence of $\nu(B=0)$~(Fig.~S7b), showing continual increase in $\nu(B=0)$ up to $T\approx 10$~K, after which the decrease in $\nu (B=0)$ can be attributed to the loss of phase coherence through thermal averaging~(Fig.~S7a). The estimated values of $T_{\mathrm{K}}$ and $T_{\mathrm{RKKY}}$ signal a competition between Kondo singlet formation and a frozen magnetic state~\cite{Doniach_Physica1977}. 
%The triangular symmetry of graphene lattice, coupled with the tendency towards antiferromagnetic alignment in the long distance limit~\cite{Saremi_PRB2007} hint towards a frustrated spin-glass state as the likely ground state. 
To gain further insight into the nature of this magnetic state, we have measured the time-averaged resistivity $\rho_{\square}$ at $n=-6\times10^{12}$~cm$^{-2}$ for the GB and SG regions of D1 simultaneously. A distinctive feature of the $T$-dependence of $\rho_{\square}$ in the GB is a noticeable downturn at $T\lesssim 1$~K (Fig.~\ref{F5}b), unlike the SG resistivity, despite accounting for quantum interference and {\it e-e} interaction corrections. Such resistivity downturn at low-$T$ is strongly indicative of spin-glass freezing resulting from reduced spin-flip scattering~\cite{Neuttiens_EPL1996,Forestier_PRB2020}. Additionally, the normalized variance $N(B=0)$ of the GB region increases rapidly by nearly an order of magnitude on cooling from $\sim 2.5$~K to $0.3$~K (Fig.~\ref{F5}c) despite the low-$T$ saturation in $\gamma_{GB}$ (Fig.~\ref{F2}e), contrasting with the behaviour of SG noise~(Fig.~S15). Such an anomalous increase cannot be explained by the standard Feng-Lee-Stone theory~\cite{Feng_PRL1986} but can be attributed to the chaotic nature of spin reorganization below the spin-glass freezing temperature~\cite{Feng_PRB1987}, as previously reported in CuMn~\cite{Israeloff_PRL1989,Fenimore_JAP1999}, AuFe~\cite{Meyer_PRB1995,Neuttiens_2000} and dilute magnetic semiconductors~\cite{Jaroszynski_PRL1998}.

In conclusion, we have identified signatures of spontaneous TRS breaking at graphene GBs using quantum transport measurements of WL and UCF. Such states emerge at high densities and at temperatures below $\sim 2$~K suggesting low-energy spin-spin interactions, possibly mediated by RKKY coupling. An anomalously sharp increase in noise below $\sim 2$~K indicates that the TRS breaking is likely due to an emergent spin glass state at graphene GBs.

%In conclusion, we have identified signatures of localized states resulting in an emergent magnetic order at graphene GBs using quantum transport measurements of WL and UCF. Spontaneous breaking of TRS at the GBs, along with a low temperature saturation in inelastic scattering rates suggest the presence of localized states at the GB which are spin-split under increased {\it e-e} interactions upon finite electron or hole doping. Spontaneous magnetization can arise at the GBs due to interactions between these local moments, possibly mediated via RKKY mechanism, but further theoretical work is required to ascertain such a possibility.

%\section{References and Notes}

%\bibliographystyle{apsrev4-2}
%\bibliography{References}


\begin{acknowledgments}
\textbf{Acknowledgements}
 We are grateful to H. R. Krishnamurthy, Sumilan Banerjee and Sudipta Dutta for useful discussions. K.H., V.K. and A.G acknowledge the Department of Science and Technology (DST) for a funded project. K.H. and A.G. also thank the National Nanofabrication Center, CeNSE, IISc (NNfC) for providing clean room facilities.

%\textbf{Author Contributions}
  %A.G. designed and supervised the project. K.H. and V.K. fabricated the devices, performed the electrical measurements and analysed the data. T.B. and M.J. provided theoretical support. A.M., G.R. and S.R. contributed materials. C.S.T. and  K.C. carried out TEM imaging and analysis. A.J. provided fabrication and experimental support. K.H., V.K., T.B., M.J. and A.G. co-wrote the manuscript. All authors discussed the results and commented on the manuscript.

%\textbf{Competing Interests}
  %The authors declare no competing financial interests.

%\textbf{Correspondence}
 %Correspondence and requests for materials should be addressed to K.H.~(email: kimberly@iisc.ac.in) or V.K.~(email: vidya@matsc.iitkgp.ac.in).
 
\end{acknowledgments}

\end{document}

\title{Spontaneous time reversal symmetry breaking at individual grain boundaries in graphene: (Supplemental Material)}

\author{Kimberly Hsieh$^{1,}$\footnote{equal contributions\label{contri}}$^{,}$\footnote{email: kimberly@iisc.ac.in}, Vidya Kochat$^{1,5,}$\textsuperscript{\ref{contri}}, Tathagata Biswas$^{1,6}$, Chandra Sekhar Tiwary$^{2,7}$, Abhishek Mishra$^3$, Gopalakrishnan Ramalingam$^4$, Aditya Jayaraman$^1$, Kamanio Chattopadhyay$^2$, Srinivasan Raghavan$^{3,4}$, Manish Jain$^1$ \& Arindam Ghosh$^{1,3}$}

\affiliation{$^1$ Department of Physics, Indian Institute of Science, Bangalore 560 012, India}

\affiliation{$^2$ Department of Materials Engineering, Indian Institute of Science, Bangalore 560 012, India}

\affiliation{$^3$ Centre for Nano Science and Engineering, Indian Institute of Science, Bangalore 560 012, India}

\affiliation{$^4$ Materials Research Center, Indian Institute of Science, Bangalore 560 012, India}

\affiliation{$^5$ Present address: Materials Science Centre, Indian Institute of Technology Kharagpur, Kharagpur 721302, India}

\affiliation{$^6$ Present address: Department of Physics, Arizona State University, Tempe, AZ 85287, USA}

\affiliation{$^7$ Present address: Department of Metallurgical and Material Engineering, Indian Institute of Technology Kharagpur, Kharagpur 721302, India}

\maketitle

\begin{lyxlist}{00.00.0000}
\item [{\textbf{S1. Methods}}]
\end{lyxlist}

\subsection{Graphene growth and device fabrication}
Graphene grains are grown by low pressure chemical vapour deposition (CVD) on Alfa Aesar (25~$\upmu$m) Cu foils. The Cu foil is wrapped to form an enclosure, and placed inside a 1~inch quartz tube reaction chamber. Initially, 100~SCCM of H$_2$ gas was passed through the reaction chamber for flushing out impurities from the Cu surface and subsequently, the Cu foil was annealed at 1000$^{\circ}$C for 4~hours while the flow rate was reduced to 50~SCCM of H$_2$. A mixture of CH$_4$, H$_2$ and N$_2$ gases in the ratio of 5:500:1000 was passed for $\sim30$~seconds following which, the furnace was cooled to room temperature under a constant flow rate of 50~SCCM of H$_2$. The Cu foil with graphene was then spin-coated with polymethyl methacrylate (MicroChem PMMA~950K) at 2000~RPM and allowed to dry at room temperature. The polymer-graphene-Cu foil stack was then suspended in 0.1~M ammonium persulphate solution to etch the Cu foil. The floating PMMA-graphene stack was rinsed thoroughly with DI water and transferred onto clean Si/SiO$_2$ (285~nm) substrates by dissolving the PMMA using hot acetone. Suitable pairs of graphene grains were chosen and patterned by electron beam lithography followed by metallization of the contacts.

\begin{figure*}[h!]
\includegraphics[width=1\linewidth]{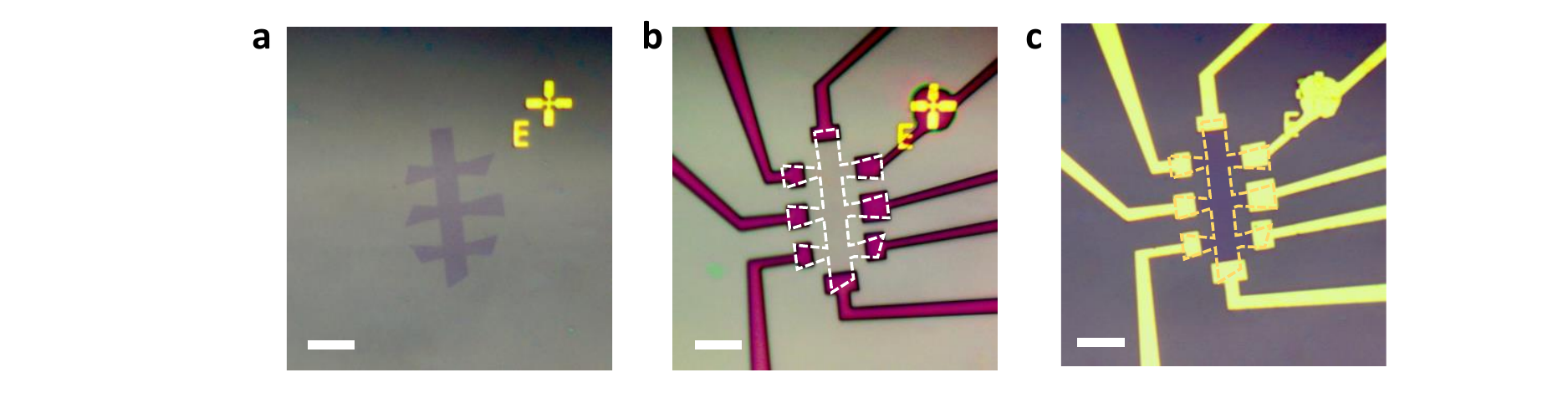}
\vspace{-0.4cm}
\caption{\textbf{Optical microscope images of the device at various stages of fabrication.} \textbf{a}~Selected pair of CVD graphene grains etched into Hall bar geometry using reactive ion etching with O$_2$ plasma. \textbf{b}~Electrical contacts defined by e-beam lithography. \textbf{c}~Final image of the device after metallization. Scale bar, 10~$\upmu$m.}
\label{Fab}
\vspace{-0.4cm}
\end{figure*}

\subsection{Transport measurements}
Magnetotransport measurements were either carried out in a Janis He-3-SSV refrigerator (for $0.3$~K~$\leq T\leq 40$~K) or a Leiden MNK-126 $^3$He/$^4$He dilution refrigerator (for $0.03$~K~$\leq T\leq 5$~K. Both time-dependent and gate voltage-dependent conductance fluctuations measurements were obtained using low-frequency lock-in method. The gate voltage-dependent UCF was measured using two-probe resistance configuration~\cite{Lee_PRB1987} where successive but overlapping windows of 4~V interval were chosen such that the conductance does not change appreciably but significant fluctuations were present for a statistically meaningful analysis ($\sim 800$ realizations). For time-dependent conductance fluctuations, an AC four-probe Wheatstone bridge configuration was used to measure the fluctuations in resistance at a fixed $n$. The time series data was digitized and decimated in multiple stages to obtain the power spectral density (PSD), which was then integrated over the experimental bandwidth to obtain the normalized variance $N=\int \frac{S_G}{G^2}df=\frac{\langle G^2\rangle}{G^2}$ (Fig.~S3a). Care was taken to minimize heating of the electrons by ensuring that the source-drain bias $V_{\mathrm{SD}} \lesssim k_{\mathrm{B}}T/e$.

\subsection{DFT calculations}
All first principles calculations have been performed within the framework of SIESTA~\cite{Soler_JPCM2002} code. We use norm-conserving pseudopotentials~\cite{Troullier_PRB1991} and Generalized Gradient approximation~\cite{Perdew_PRL1996} (PBE) for exchange correlation functional. A double-$\zeta$ plus polarization (DZP) basis set and a mesh cut-off of 300~Ry have been chosen for Brillouin zone integrations. For the calculations of both GB$(2,0)|(2,0)$ and GB$(5,0)|(3,3)$, $1\times10\times50$ k-grid have been employed. All the atomic positions are relaxed using conjugate-gradient algorithm until the forces were less than 0.04~eV$/$\AA.

\newpage

\begin{lyxlist}{00.00.0000}
\item [{\textbf{S2. Weak localization measurements for D1 at \boldmath{$T=0.3$}~K}}]
\end{lyxlist}

\begin{figure*}[h!]
\includegraphics[width=1\linewidth]{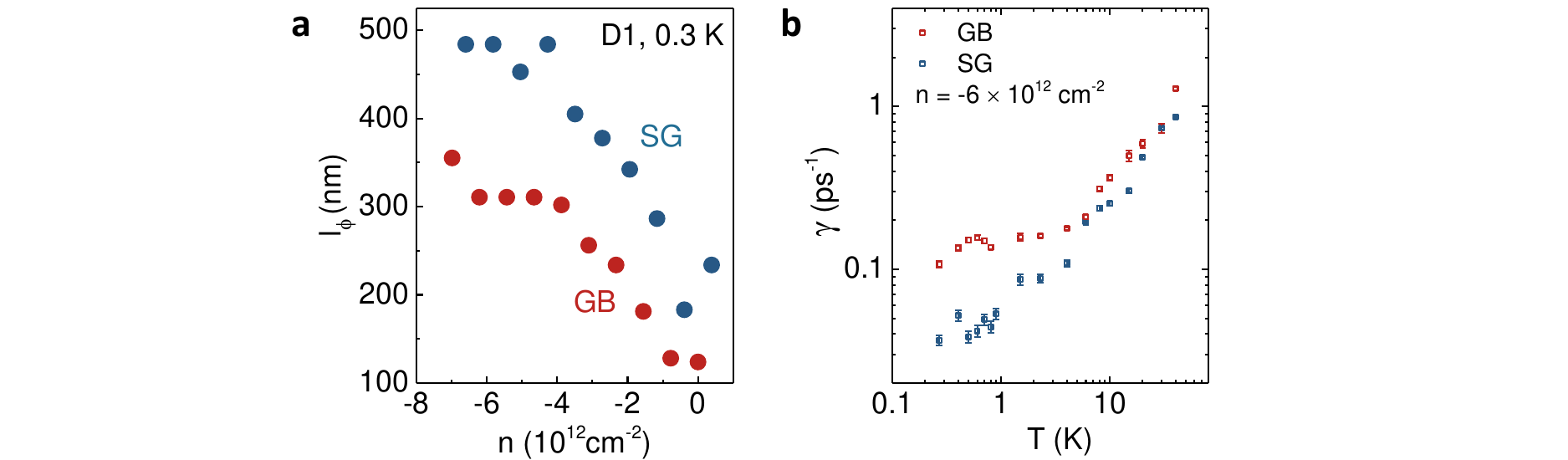}
\vspace{-0.4cm}
\caption{\textbf{a},~The $l_{\phi}$ values at 0.3~K obtained from the magnetoresistance fitting for graphene is plotted for the SG and GB regions as a function of $n$. \textbf{b},~~The scattering rates $\gamma$ (obtained from the HLN fits to the magnetoconductance data) plotted as a function of temperature $T$ for the SG and GB regions at $n=-6\times10^{12}$~cm$^{-2}$.} %\textbf{c},~The spin flip scattering rate $\gamma_s$ for the GB region is plotted as a function of $T$, exhibiting a maximum at $T_{\mathrm{K}}\approx 0.5$~K.}
\label{lphi}
\vspace{-0.4cm}
\end{figure*}

%The Kondo temperature $T_{\mathrm{K}}$ can be extracted from the temperature dependence of the spin flip rate $\gamma_s$ by the Nagaoka-Suhl relation~\cite{Pierre_PRB2003,Suhl_1973}

%\begin{equation}
%\gamma_s=\frac{c_{mag}}{\pi\hbar N(E_{\mathrm{F}})}\frac{\pi^2 S(S+1)}{\pi^2 S(S+1)+\ln^2(T/T_{\mathrm{K}})}
%\end{equation}

%\noindent where $c_{mag}$ is the magnetic impurity density, $S=1/2$ and $N(E_{\mathrm{F}})$ is the density of states of the host material at the Fermi level.

\begin{lyxlist}{00.00.0000}
\item [{\textbf{S3. Noise for D1 at \boldmath{$B=0$}~T}}]
\end{lyxlist}

\begin{figure*}[h!]
\includegraphics[width=1\linewidth]{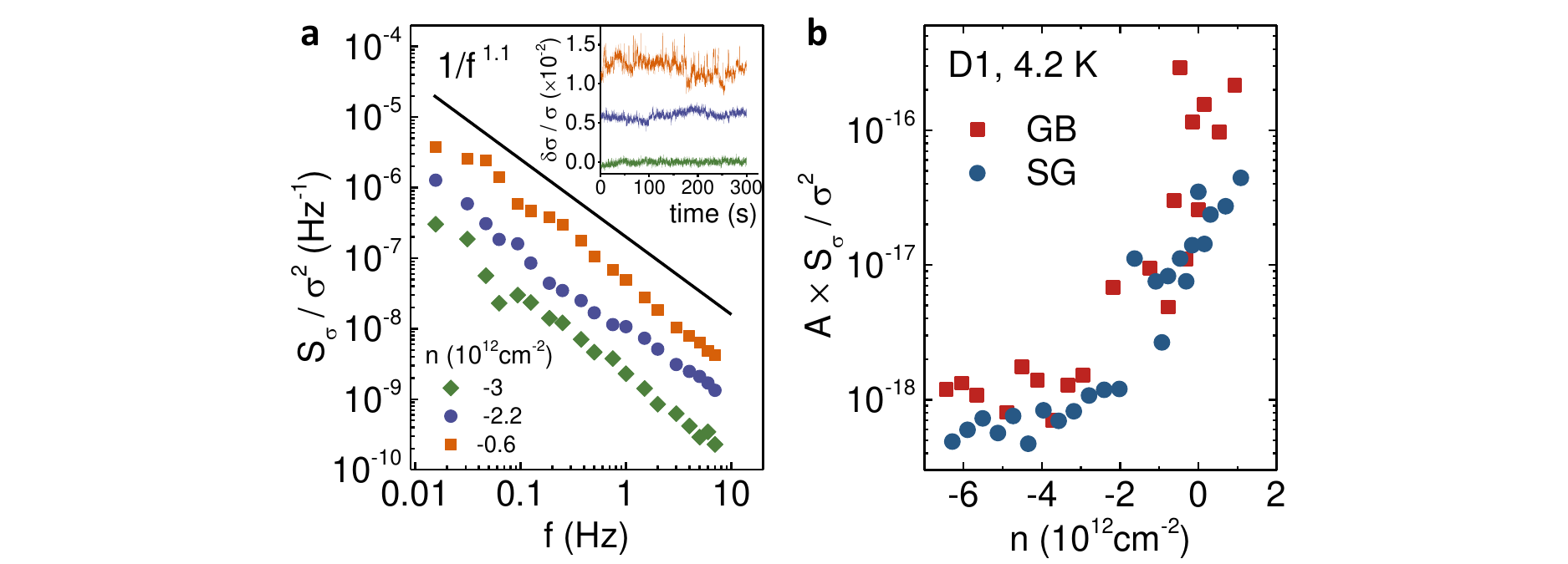}
\vspace{-0.4cm}
\caption{\textbf{\boldmath{$1/f$} noise for D1 at zero magnetic field.} \textbf{a},~The normalized power spectrum for D1 at $T=4.2$~K is plotted for three different $n$ showing clear $1/f$ behaviour. The inset shows the conductivity fluctuations for these $n$ clearly showing that the fluctuation magnitude is increased at low $n$. \textbf{b},~The area-normalized power spectral density~($A\times{S_\sigma/\sigma^2}$) is shown for the SG and GB regions of D1 as a function of $n$ at $T=4.2$~K.} %\textbf{c},~${S_\sigma/\sigma^2}$ as a function of temperature is plotted for the GB region at $n=-6.9\times10^{12}$~cm$^{-2}$ clearly indicating $1/T$ dependence at lower temperatures.
\label{Noise_D1}
\vspace{-0.4cm}
\end{figure*}

The time-dependent conductance fluctuations in the inter-grain region measured at different carrier densities is shown in the inset of Fig.~\ref{Noise_D1}a. The power spectrum obtained from these conductivity fluctuations exhibit $1/f$ behaviour in the frequency
domain as shown in Fig.~\ref{Noise_D1}a.  The area-normalized power spectrum~($A\times{S_\sigma/\sigma^2}$) as a function of $n$ at 4.2~K is shown
in Fig.~\ref{Noise_D1}b for the SG and GB regions. The noise in the GB region is larger than the SG region indicating more dynamic scattering arising from the localized states at the GBs~\cite{Vidya_Nanoletters2016}. 
%The noise $S_\sigma/\sigma^2$ in the inter-grain region at $n=-6.9\times10^{12}$~cm$^{-2}$ (Fig.~\ref{Noise_D1}c) shows a clear non-monotonic temperature dependence. The noise initially decreases as the temperature is lowered in an almost linear fashion~($\propto{T}$) as expected from the noise models based on carrier tunneling between the channel and trap states spread over an energy range $\sim k_\mathrm{B} T$ about the Fermi energy.  Below a certain temperature, the noise increases with further decrease in $T$ with a $1/T$ behaviour which can be explained using the Feng-Lee-Stone~(FLS) theory of UCF noise~\cite{Feng_PRL1986}. According to FLS theory, $S_\sigma/\sigma^2\propto{n_T}l_\phi^4$, with $l_\phi\propto1/\sqrt{T}$ and the density of active two level systems, $n_T\propto{T}$ which gives the observed $1/T$ behaviour, clearly indicating that universal conductance fluctuations is the dominant noise mechanism.

\begin{lyxlist}{00.00.0000}
\item [{\textbf{S4. UCF from time-dependent conductance fluctuations}}]
\end{lyxlist}

For time-dependent conductance fluctuations, an AC four-probe Wheatstone bridge configuration was used to measure the fluctuations in resistance at a fixed $n$ and a fixed $B$. Two resistors, $R1$ and $R2$ form the upper arms of the Wheatstone's bridge. $R1\sim 1$~M$\Omega$ is the current limiting resistor through which the sample is biased using an AC voltage from the lock-in amplifier (SR830). The variable resistor $R3$ in the third arm is designed using a series of low drift 1~k$\Omega$, 10~k$\Omega$ and 100~k$\Omega$ potentiometers. For each $B$ at a fixed $V_{\mathrm{BG}}$, the bridge is balanced by varying $R3$ and the small unbalanced voltage is first amplified using the SR560 preamplifier with a chosen gain of 100 and then measured in the A-B mode of the lock-in amplifier with a high sensitivity, with the B port connected to ground via a 50~$\Omega$ terminator. The entire setup is shielded using a grounded Faraday cage to minimize electromagnetic interference. The signal from both the in-phase (X) and quadrature (Y) channels of the lock-in amplifier are digitized at a high sampling frequency $f_s$ (1000 samples per second) using a 16-bit ADCs (analog-to-digital converters) NI USB-6210 from National Instruments with a maximum sampling rate of 250~Kilosamples per second, 4095 samples of on-board memory and an input impedance of $\sim 10$~G$\Omega$, which can be interfaced with LabVIEW programs. The data is stored in the on-board memory temporarily and then transferred to the hard disk of the computer in segments for further processing.

The time series data was then decimated in multiple stages to obtain the power spectral density (PSD) using the fast Fourier transform (FFT) technique, which was then integrated over the experimental bandwidth to obtain the normalized variance $N=\int \frac{S_G}{G^2}df=\frac{\langle G^2\rangle}{G^2}$ at each $B$ for a fixed $V_{\mathrm{BG}}$. Care was taken to minimize heating of the electrons by ensuring that the source-drain bias $V_{\mathrm{SD}} \lesssim k_{\mathrm{B}}T/e$. More details of the experimental technique can be found in Ref.~\cite{Ghosh_arXiv2004}. The evaluation of the variance $\langle\delta G_{\phi}^2\rangle$ was obtained from statistically meaningful windows of time ($\approx 1200$~seconds with $\approx 18750$ realizations of $G$).

\begin{lyxlist}{00.00.0000}
\item [{\textbf{S5. UCF from gate voltage-dependent conductance fluctuations}}]
\end{lyxlist}

\begin{figure*}[b!]
%\begin{center}
\includegraphics[width=1\linewidth]{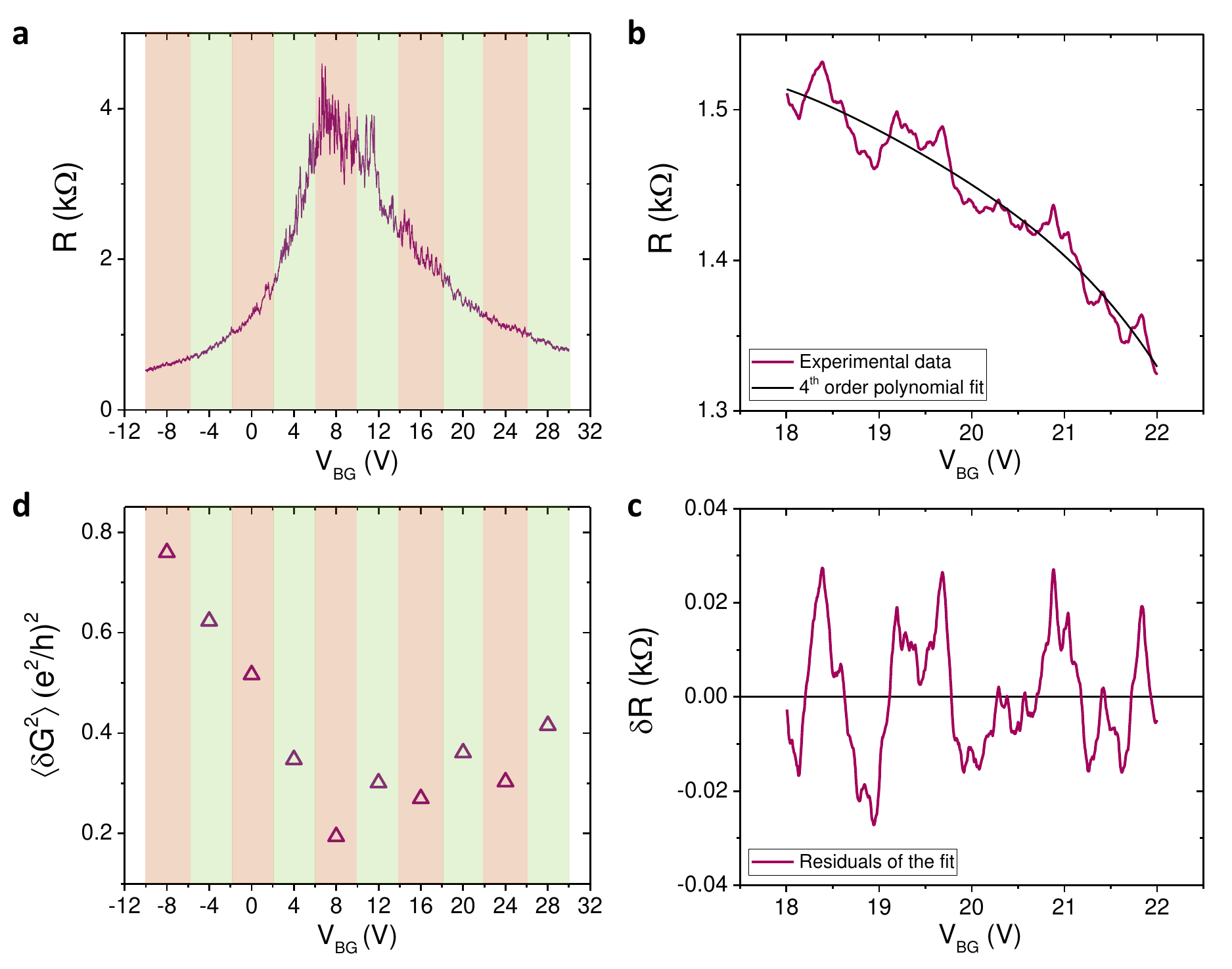}
\caption{\label{UCFEvaluation} \textbf{a},~A typical plot of two-probe resistance as a function of back gate voltage for graphene, split into successive 4~V interval windows. \textbf{b},~The resistance fluctuations within a 4~V window range plotted along with its fourth order polynomial fit. \textbf{c}~The residuals obtained from the fourth order polynomial fit of the curve in \textbf{b}. \textbf{d}~The calculated values of $\langle \delta G^2\rangle$ corresponding to the different gate voltage windows of the representative curve in \textbf{a}.}
%\end{center}
\end{figure*}

UCF measurements were performed in D2 using the two-probe resistance configuration~\cite{Lee_PRB1987}. The phase coherence length ($l_{\phi}$) was extracted from the WL MC curves using the modified Hikami-Larkin-Nagaoka expression~\cite{McCann_PRL2006}. Successive gate voltage windows of 4~V interval were chosen such that the conductance does not change appreciably but significant fluctuations are nevertheless present for a statistically meaningful analysis ($\sim 800$ realizations). This is demonstrated in Fig.~\ref{UCFEvaluation}a for a typical transfer characteristic curve of graphene. The two-probe resistance in each $V_{\mathrm{BG}}$ window was then fitted with a fourth-order polynomial as shown in Fig.~\ref{UCFEvaluation}b, and then the variance of the fluctuations was calculated from the residuals of the fit as shown in Fig.~\ref{UCFEvaluation}c. The fitting with a fourth-order polynomial was done to subtract a smooth background for calculating the fluctuations and to suppress slowly varying changes in the conductance. Fitting with polynomials of other orders do not give any qualitative difference in the analysis~\cite{Pal_PRL2012,Vidya2014}.  Finally, the magnitude of conductance fluctuations is obtained from the resistance fluctuations as

\begin{equation}\label{RFtoCF}
  \langle \delta G^2\rangle=\frac{\langle \delta R^2\rangle}{\langle R\rangle^4}
\end{equation}

\noindent where $\langle R\rangle$ is the average four-terminal resistance corresponding to that particular gate voltage window. However, this measured $\langle \delta G^2\rangle$ arises from the entire sample and in order to obtain the magnitude of conductance fluctuations within a phase-coherent box of area $l_{\phi}\times l_{\phi}$, the principle of superposition is invoked which gives

\begin{equation}\label{UCF_Superposition}
  \frac{\langle \delta G^2\rangle}{\langle G\rangle^2}=\frac{1}{N}\frac{\langle \delta G_{\phi}^2\rangle}{\langle G_{\phi}\rangle^2}
\end{equation}

\noindent where $N=LW/l_{\phi}^2$ is the number of phase coherent boxes within the channel of length $L$ and width $W$. Since $\langle \delta G_{\phi}^2\rangle=\sigma$ and $\langle \delta G^2\rangle=\sigma W/L$,

\begin{equation}\label{UCF_PhaseCoherent}
  \langle \delta G_{\phi}^2\rangle=\frac{L^3}{W}\frac{\langle \delta G^2\rangle}{l_{\phi}^2}
\end{equation}

\noindent Both the phase breaking length ($l_{\phi}$) and the variance of conductance fluctuations ($\langle \delta G^2\rangle$) are dependent on carrier density and have to be evaluated experimentally to extract the density dependence of $\langle \delta G_{\phi}^2\rangle$.
\newpage

\begin{lyxlist}{00.00.0000}
\item [{\textbf{S6. Magneto-noise data for devices D2 and D3}}]
\end{lyxlist}

\setcounter{subsection}{0}

\subsection{D2 at \boldmath{$T = 0.3$}~K}  

\begin{figure*}[h!]
%\begin{center}
\includegraphics[width=1\linewidth]{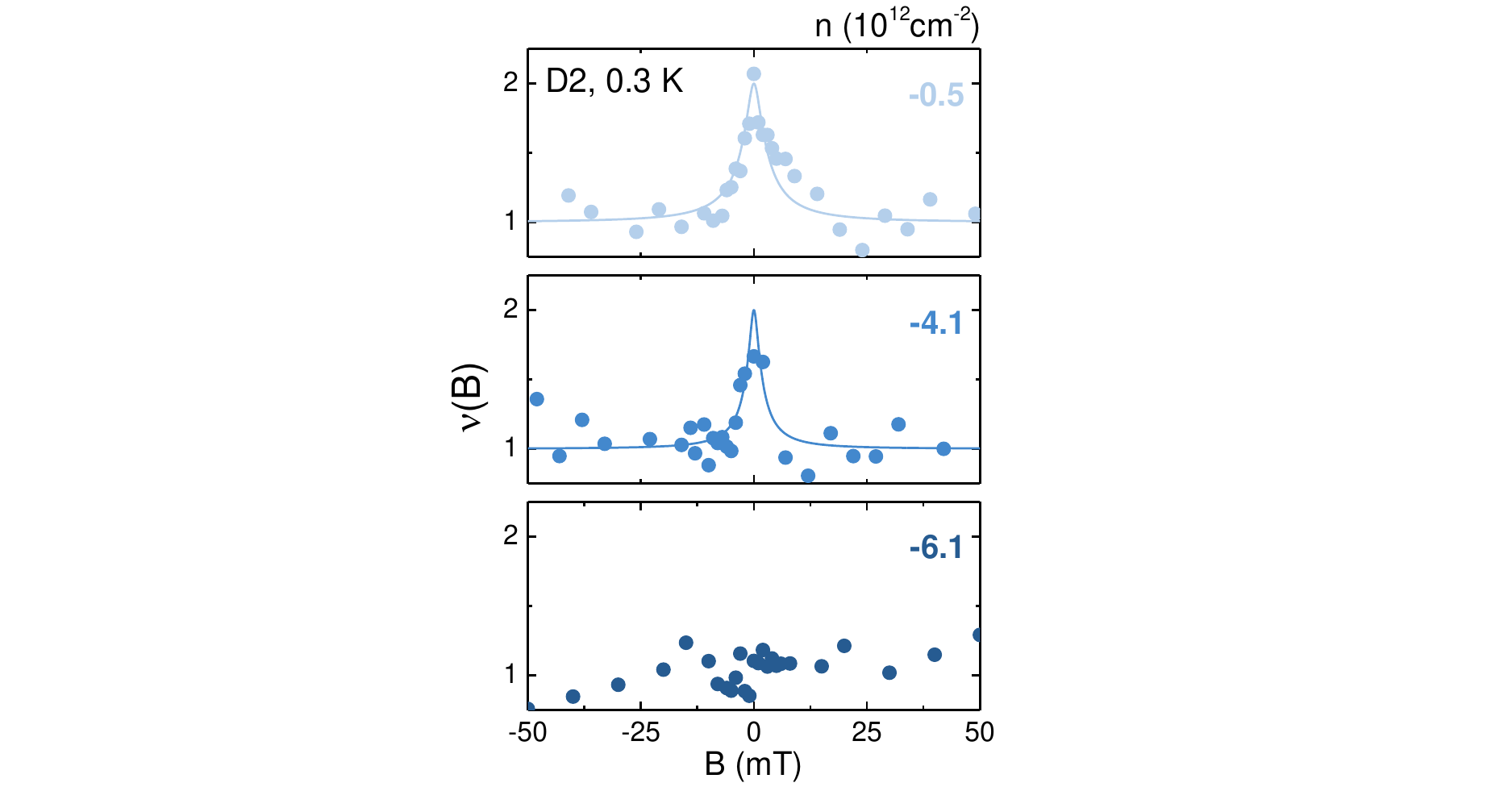}
\caption{\label{D2_285mK} $\nu(B)$ for D2 at $T=0.3$~K for different carrier densities $n$ in units of $10^{12}$~cm$^{-2}$. Solid lines are fits to the crossover function.}
%\end{center}
\end{figure*}

\newpage

\subsection{D2 at \boldmath{$T = 4.5$}~K}

\begin{figure*}[h!]
%\begin{center}
\includegraphics[width=1\linewidth]{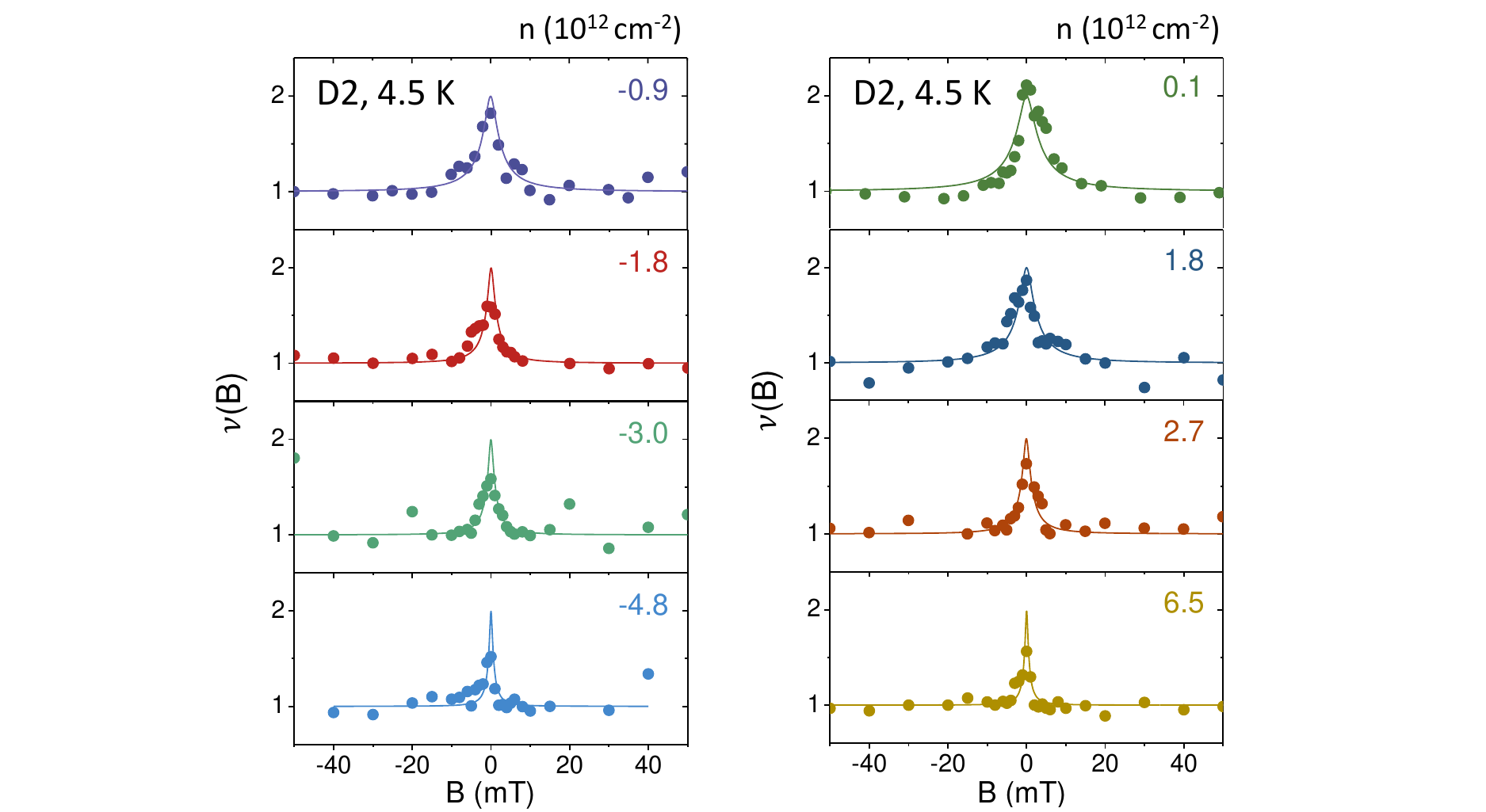}
\caption{\label{D2_4p5K} $\nu(B)$ for D2 at $T=4.5$~K for different carrier densities $n$ in units of $10^{12}$~cm$^{-2}$. Solid lines are fits to the crossover function.}
%\end{center}
\end{figure*}

\newpage 
\subsection{D2 at higher \boldmath{$T$}}

\begin{figure*}[h!]
\includegraphics[width=1\linewidth]{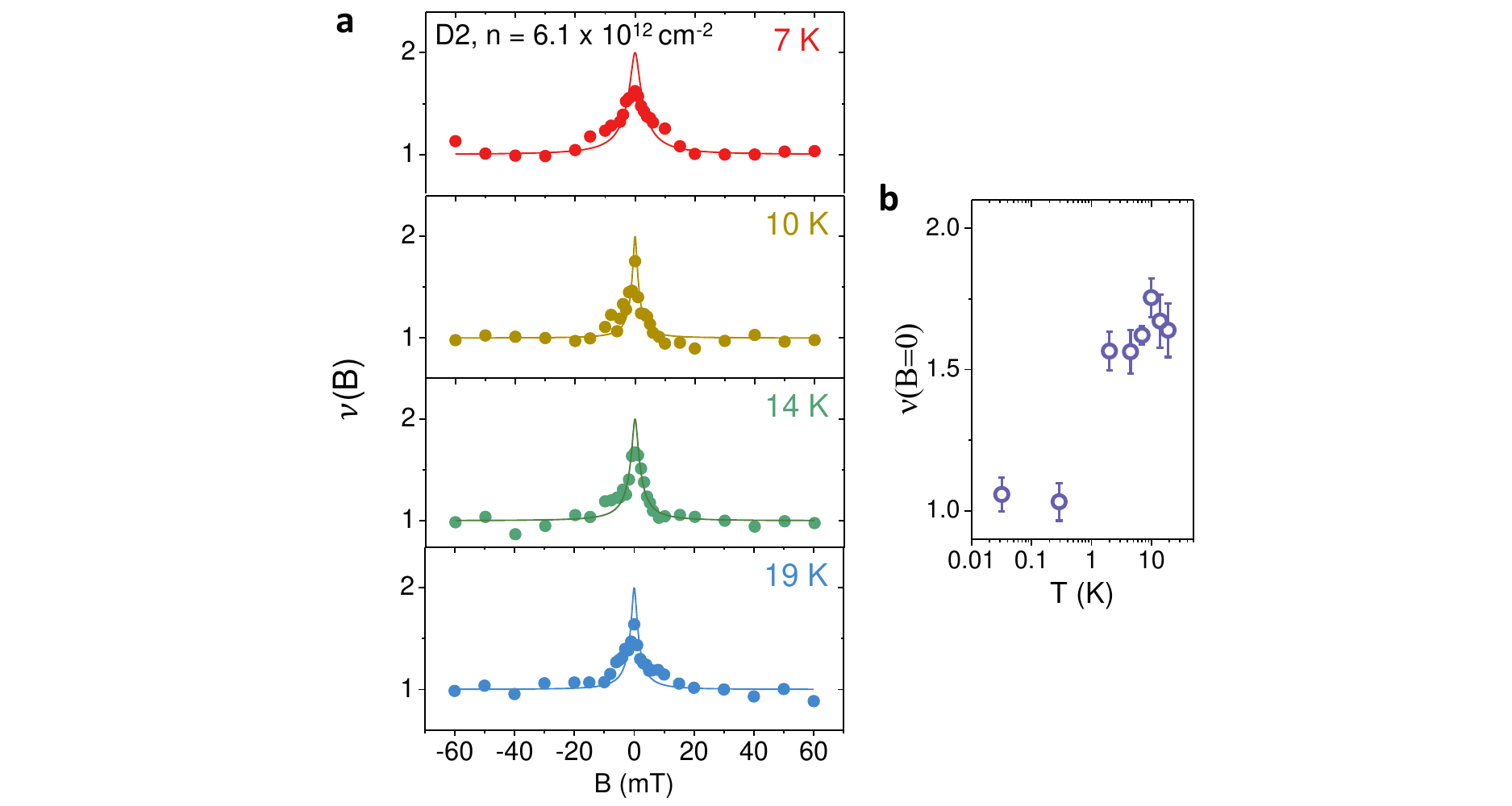}
\caption{\label{D2_highT} \textbf{a},~$\nu(B)$ for GB of D2 at $T=7$, $10$, $14$ and $19$~K for $n=6.1\times 10^{12}$~cm$^{-2}$. Solid lines are fits to the crossover function. \textbf{b},~The reduction factor $\nu(B=0)$ of the time-dependent UCF in the intergrain region plotted as a function of temperature $T$ at high electron density $n=6.1\times10^{12}$~cm$^{-2}$.}
\end{figure*}

From Fig.~\ref{D2_highT}, we observe that the spontaneous TRS breaking in graphene GBs happens only at ultra-low temperatures below $\sim 1-2$~K. $\nu(B=0)$ continues to increase as $T$ increases up to $\approx 10$~K, after which $\nu (B=0)$ starts to decrease as quantum interference effects begin to disappear due to loss of phase coherence. At higher temperatures, the $1/f$ noise can no longer be attributed entirely to UCF~\cite{Trionfi_PRB2004}.

\newpage
\subsection{D3 at \boldmath{$T = 8$}~K}

\begin{figure*}[h!]
%\begin{center}
\includegraphics[width=1\linewidth]{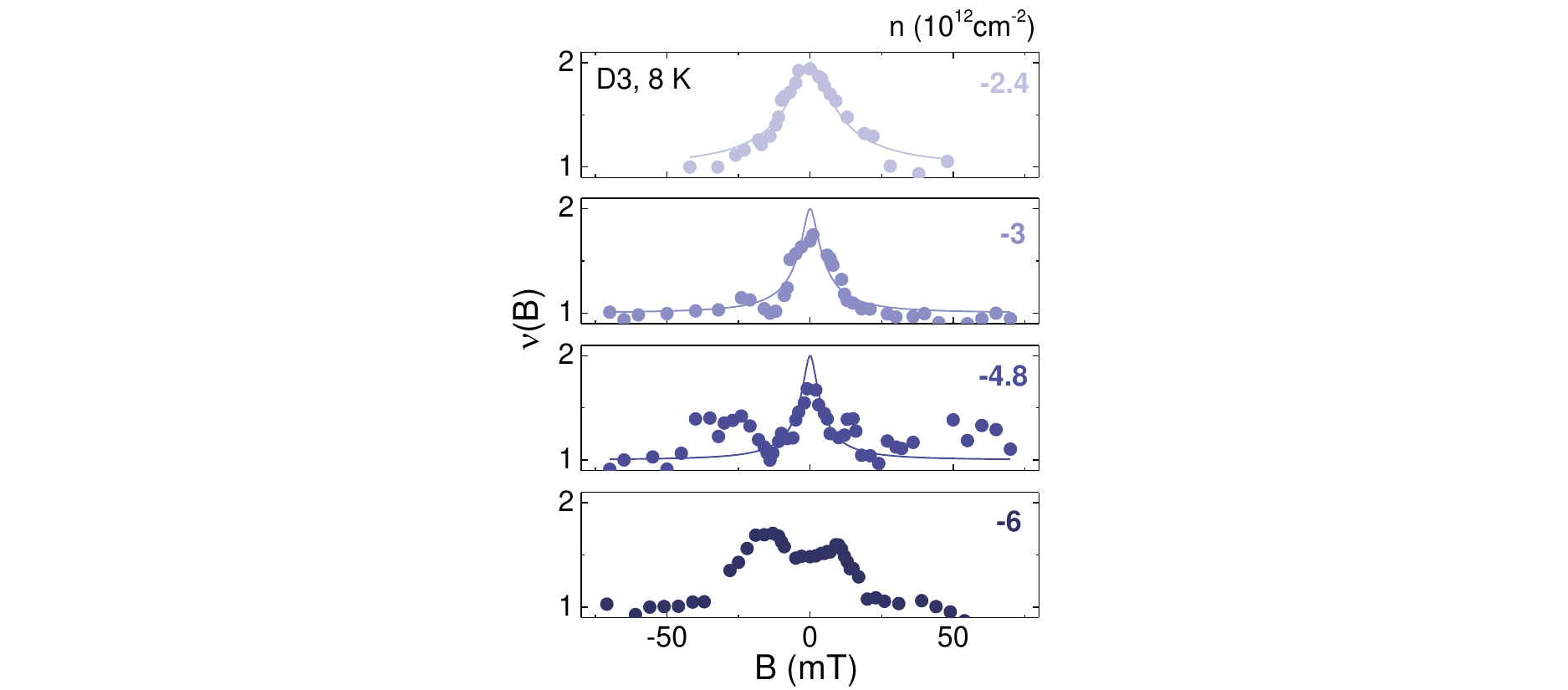}
\caption{\label{D3_8K} $\nu(B)$ for D3 at $T=8$~K for different carrier densities $n$ in units of $10^{12}$~cm$^{-2}$. Solid lines are fits to the crossover function.}
%\end{center}
\end{figure*}

\newpage
\subsection{Single grain magneto-noise for D2 at \boldmath{$T = 4.5$}~K}

\begin{figure*}[h!]
%\begin{center}
\includegraphics[width=1\linewidth]{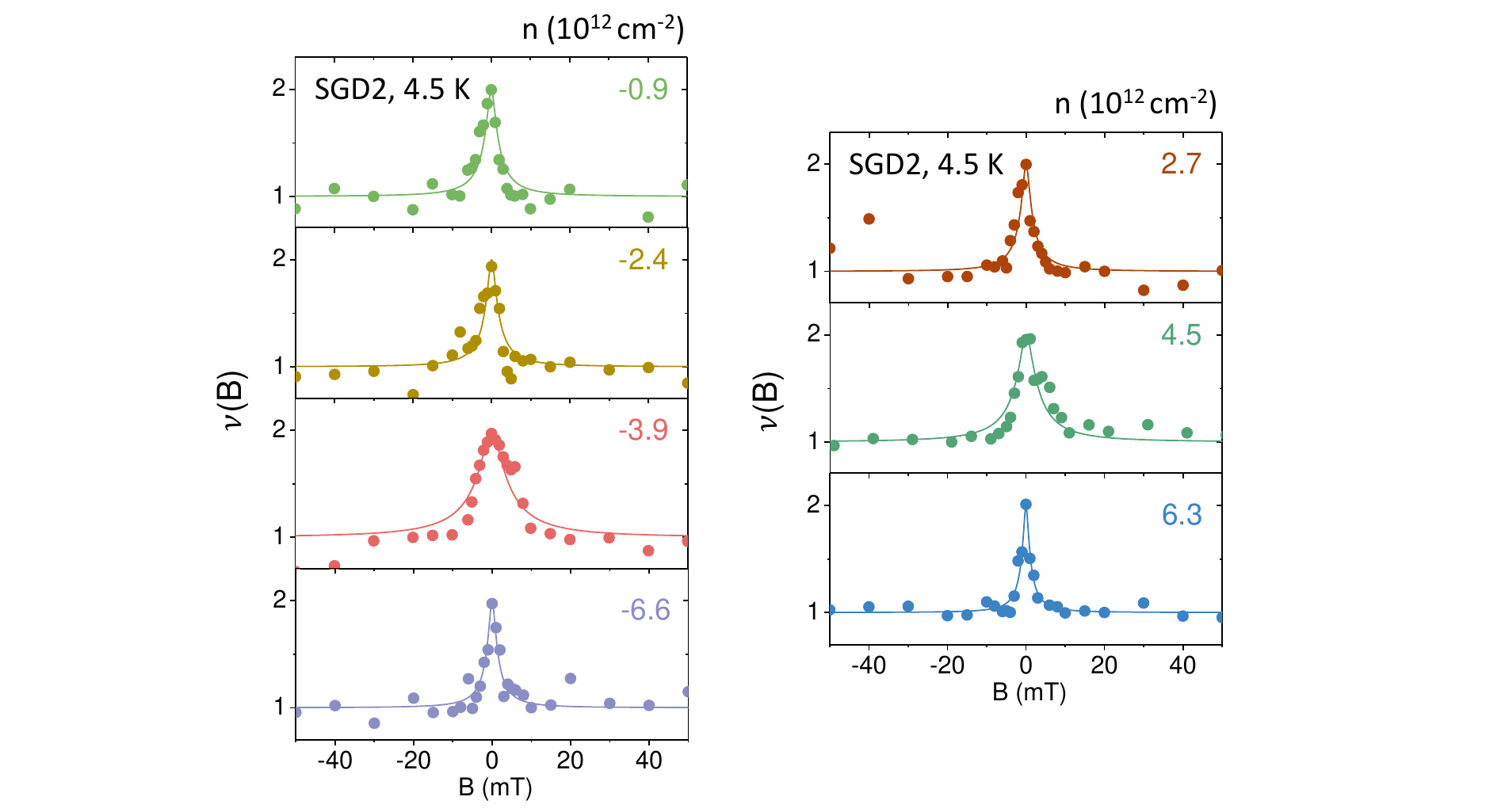}
\caption{\label{D2_SG_4p5K} $\nu(B)$ for the single grain region (without GB) of D2 at $T=4.5$~K for different carrier densities $n$ in units of $10^{12}$~cm$^{-2}$, showing factor-of-two reduction in the magneto-noise even at high $n$ unlike the case of the inter-grain region. Solid lines are fits to the crossover function.}
%\end{center}
\end{figure*}

\newpage
\begin{lyxlist}{00.00.0000}
\item [{\textbf{S7. Density functional calculations for atomically sharp grain boundaries}}]
\end{lyxlist}

\begin{figure}[!h]
\begin{center}
\includegraphics [width=1\linewidth]{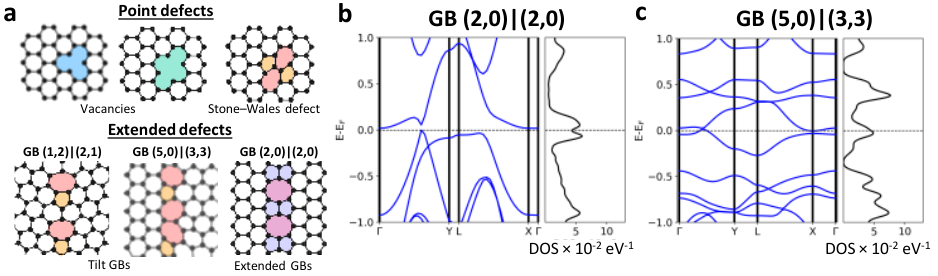}
\end{center}
\vspace{-0.4cm}
\caption{\textbf{a},~Schematic showing two categories of defects in graphene: point and extended defects. \textbf{b,c},~Bandstructure and density of states calculations for graphene with GB containing \textbf{b},~pentagon-octagon defects (GB$(2,0)|(2,0)$) and \textbf{c},~pentagon-heptagon defects (GB$(5,0)|(3,3)$) showing enhanced DOS due to flat band formation.}
\label{GB_DFT}
\vspace{-0.4cm}
\end{figure}

\begin{lyxlist}{00.00.0000}
\item [{\textbf{S8. Density functional calculations for a grain boundary of finite width}}]
\end{lyxlist}

GB$(6,6)|(6,6)$ can be thought of as a clean translational GB$(2,0)|(2,0)$ sandwiched between two GB$(3,3)|(5,0)$, type of GBs~\cite{Alexandre_NanoLett2012}. GB$(3,3)|(5,0)$ consists of only Stone-Wales 5-7 defects while GB$(2,0)|(2,0)$ contains only 5-8 defect pairs. To match the periodicity in the direction parallel to the GB, five unit cells of GB$(2,0)|(2,0)$ and two unit cells of GB$(3,3)|(5,0)$ were used resulting in a matching vector (10,0) in region between the two GBs and (6,6) in the region beyond either GB$(3,3)|(5,0)$. For the calculations of GB$(2,0)|(2,0)$,  $1\times10\times50$ k-grid was employed. However, for the newly proposed GB$(6,6)|(6,6)$, the number of k-points along the z-direction was reduced by a factor of 5 due to the larger supercell size. The calculations were carried out on a supercell of 242~atoms where the separation between two adjacent GBs in GB$(6,6)|(6,6)$ was fixed at $\sim10$~\AA. 

\begin{figure*}[t!]
%\begin{center}
\includegraphics[width=1\linewidth]{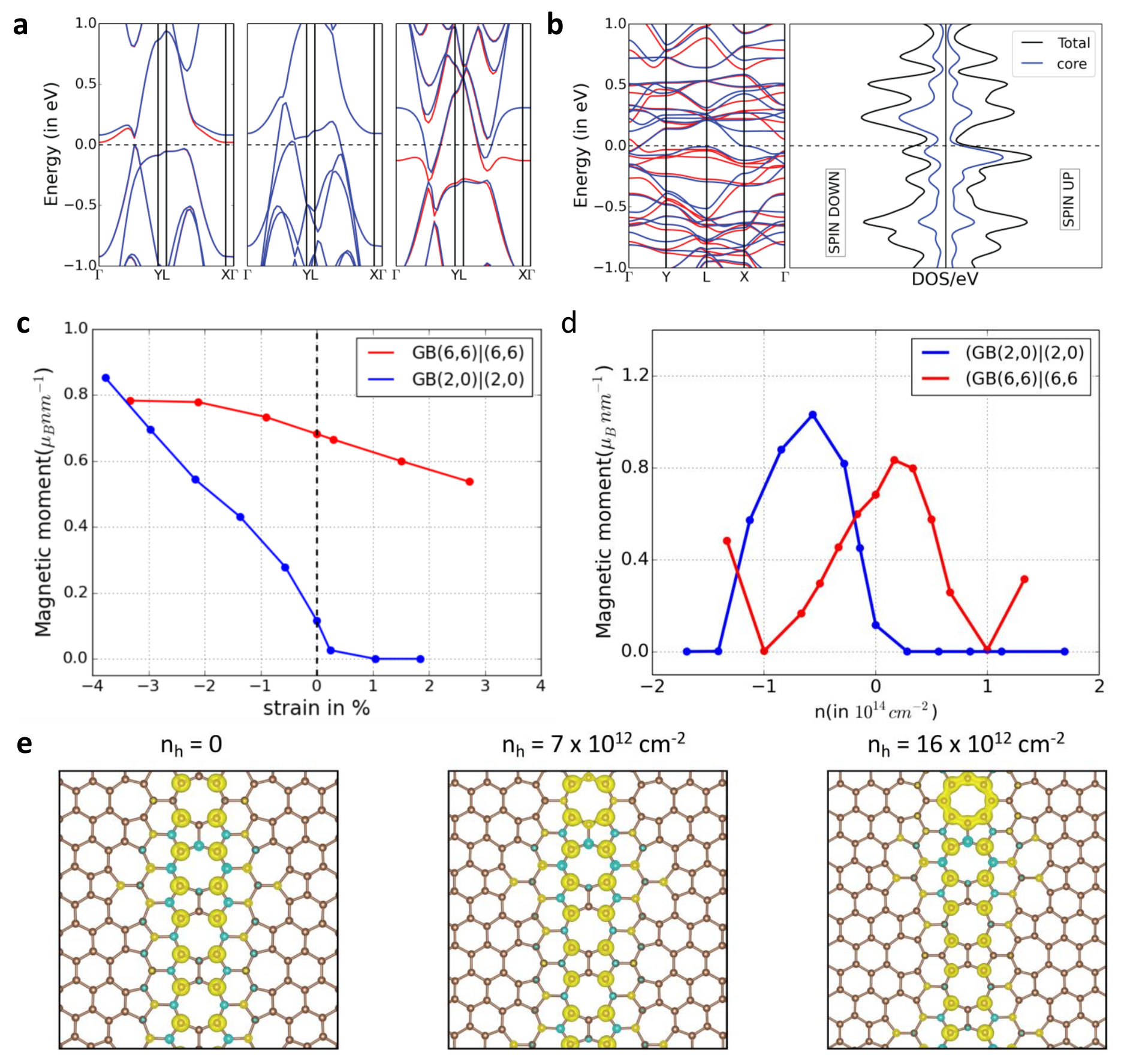}
\caption{\label{DFT} \textbf{a},~Band structure of GB$(2,0)|(2,0)$ where the three panels represent zero doping, hole doping of $-5.6\times10^{13}$~cm$^{-2}$ and electron doping of $5.6\times10^{13}$~cm$^{-2}$ respectively. \textbf{b},~Left panel shows the band structure of and spin polarized density of states (DOS) for GB$(6,6)|(6,6)$. Right panel shows the projected DOS for only the core atoms. \textbf{c,d},~Magnetic moment per unit length of the GB for both GB$(2,0)|(2,0)$ and GB$(6,6)|(6,6)$ as a function of \textbf{c} uniaxial strain along the armchair direction and \textbf{d} carrier density $n$ . \textbf{e}~Isosurface showing the spin polarization density along the GB$(6,6)|(6,6)$ at different densities of hole doping.}
%\end{center}
\end{figure*}

Fig.~\ref{DFT}a shows the band structure of GB$(2,0)|(2,0)$ for three cases of zero-, hole- and electron-doping. The ferromagnetic state arises only for the case of electron doping as the Fermi level is raised due to Stoner instability. The localized core states in the GB regions give rise to flat bands near the Fermi level which split near the $\Gamma$-point resulting in a non-zero magnetic moment. On the other hand, the band structure and density of state (DOS) plots for GB$(6,6)|(6,6)$ show spin-splitting not only near the $\Gamma$-point but throughout the entire Brillouin zone. The origin of the magnetic moments may be attributed to the in-built lattice strain due to the additional 5-7 defects around the GB region. It is important to note that GB$(6,6)|(6,6)$ has a lattice dimension five times larger than that of GB$(2,0)|(2,0)$ in the Y-direction, hence it is necessary to take zone folding into account while comparing the band structures of the two GBs. The additional flat bands near the Fermi level may come directly from the random distribution of Stone-Wales defects or via internal strain due to their presence. The spin-split bands are not only spread throughout the Brillouin zone but also across an energy range. The bands are split in the range of 1~eV above and below the Fermi level, which explains the dependence of the magnetic moment per unit length as a function of changing carrier density (Fig.~\ref{DFT}d). The doping response can also be explained from the spin polarized DOS plots for GB$(6,6)|(6,6)$ shown in the right panel of Fig.~\ref{DFT}b. The core atoms consist of the 5-8 defects which host almost all the magnetic moments as seen in the isosurface plots of Fig.~\ref{DFT}e.

Consistent with the findings of previous theoretical studies~\cite{Alexandre_NanoLett2012,Kou_ACSNano2011,Cortijo_NuclPhysB2007}, 5-8 defects of translational GB$(2,0)|(2,0)$ were found to host a ferromagnetic ground state when the system is doped~\cite{Alexandre_NanoLett2012} or strained~\cite{Kou_ACSNano2011}. Our theoretical calculations show that application of a tensile strain along the zigzag direction or compressive strain along the armchair direction significantly enhances the magnetic moment per unit length along the GB (Fig.~\ref{DFT}c). The strain is believed to arise during the growth and subsequent transfer process of the graphene grains on to SiO$_2$ substrate. Fig.~\ref{DFT}c also reveals a striking difference between the two GBs. GB$(6,6)|(6,6)$ exhibits a significantly large magnetic moment even in the absence of external strain. The response to an external strain along the armchair direction is much lesser for GB$(6,6)|(6,6)$ as compared to GB$(2,0)|(2,0)$ since the additional 5-7 defects around the core 5-8 defects in GB$(6,6)|(6,6)$ are expected to absorb a significant fraction of the strain applied.

The variation of magnetic moments as a function of carrier density $n$ is plotted in Fig.~\ref{DFT}d. For GB$(2,0)|(2,0)$, electron doping induces significant magnetic moment formation which diminishes once the doping density is increased above $\sim -0.5\times 10^{14}$~cm$^{-2}$, far beyond the range accessible by an SiO$_2$ back gate. This can be explained from the sharp resonance in the DOS located at about $\pm0.1$~eV of the Fermi level, which consists of the localized core states in the GB region. The magnetic moments originate from the Coulomb interaction between the electrons in these states~\cite{Alexandre_NanoLett2012}. In addition to the fact that there is a significantly large magnetic moment formation at zero doping for GB$(6,6)|(6,6)$, the response to doping is also very different from that of GB$(2,0)|(2,0)$. The magnetic moments are formed for both hole and electron doping unlike the case of GB$(2,0)|(2,0)$ as explained earlier from the band structure and DOS shown in Fig.~\ref{DFT}a. The isosurface of spin polarization density is showing for GB$(6,6)|(6,6)$ in Fig.~\ref{DFT}e for the undoped system as well as two different hole doping densities. The spin polarization value was chosen as $1.5\times10^{-3}$~$\mu_B$ for plotting these isosurfaces. For the undoped system, the magnetic moments are localized on the atoms along the zigzag edges in the middle of the GB, consistent with the case of GB$(2,0)|(2,0)$~\cite{Kou_ACSNano2011,Alexandre_NanoLett2012}. As the number density is tuned, the C-C dimer connecting the two zigzag edges begin to host magnetic moments as well. On increasing the hole density further, some of the atoms of the different sublattice of the zigzag edge reverse their spin polarizations which leads to a net enhancement of the overall magnetic moment of the GB. These moments may explain the spontaneous breaking of time reversal symmetry at higher densities, if they interact to result in a frozen magnetic order.

\begin{lyxlist}{00.00.0000}
\item [{\textbf{S9. DOS per site for different types of GBs}}]
\end{lyxlist}

\begin{figure*}[h!]
%\begin{center}
\includegraphics[width=1\linewidth]{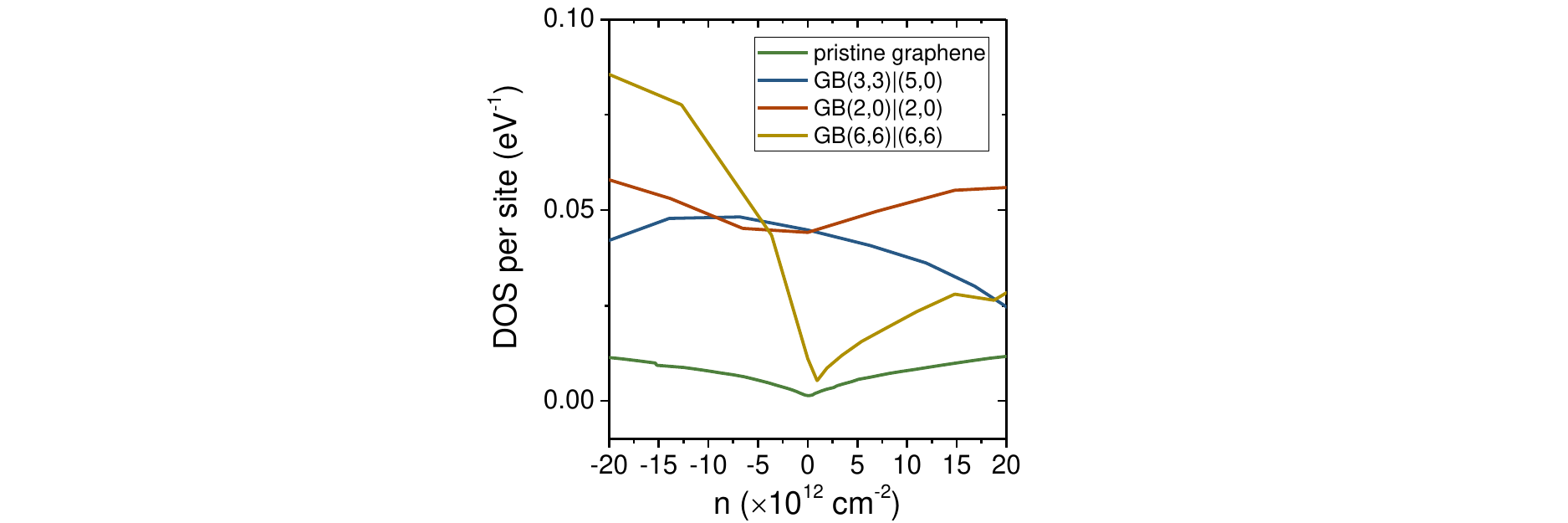}
\caption{\label{DOSvsn} DOS per site as a function of $n$ for pristine graphene (green), GB$(3,3)|(5,0)$ with only pentagon-heptagon defects (blue), GB$(2,0)|(2,0)$ with only pentagon-octagon defects (red) and finite width GB$(6,6)|(6,6)$ with multiple types of defects (yellow).}
%\end{center}
\end{figure*}

\newpage

\begin{lyxlist}{00.00.0000}
\item [{\textbf{S10. Mapping the defect disorder in a grain boundary of finite width}}]
\end{lyxlist}

\begin{figure*}[h!]
%\begin{center}
\includegraphics[width=1\linewidth]{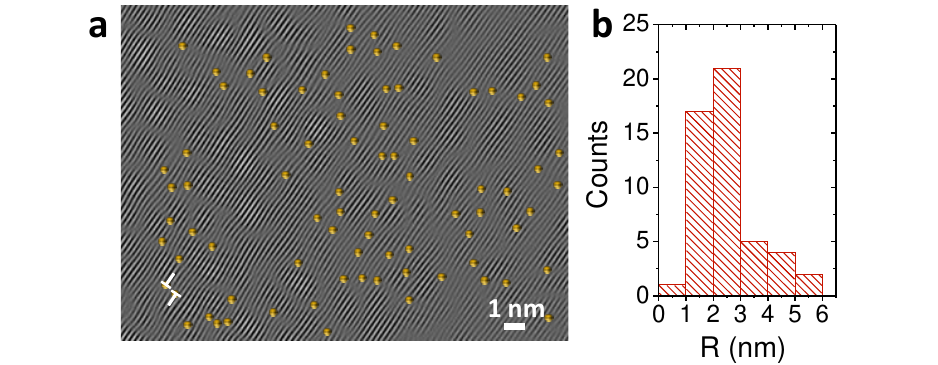}
\caption{\label{TEM} \textbf{a},~Experimental demonstration of defect-induced distortion on the parallel lattice array of a representative HRTEM image of a finite-width GB. \textbf{b},~Histogram obtained from the HRTEM image in \textbf{a} revealing an average defect distance $R\approx 2$~nm.}
%\end{center}
\end{figure*}

\begin{lyxlist}{00.00.0000}
\item [{\textbf{S11. Estimation of Kondo temperature \boldmath{$T_{\mathrm{K}}$}}}]
\end{lyxlist}

\begin{figure*}[h!]
%\begin{center}
\includegraphics[width=1\linewidth]{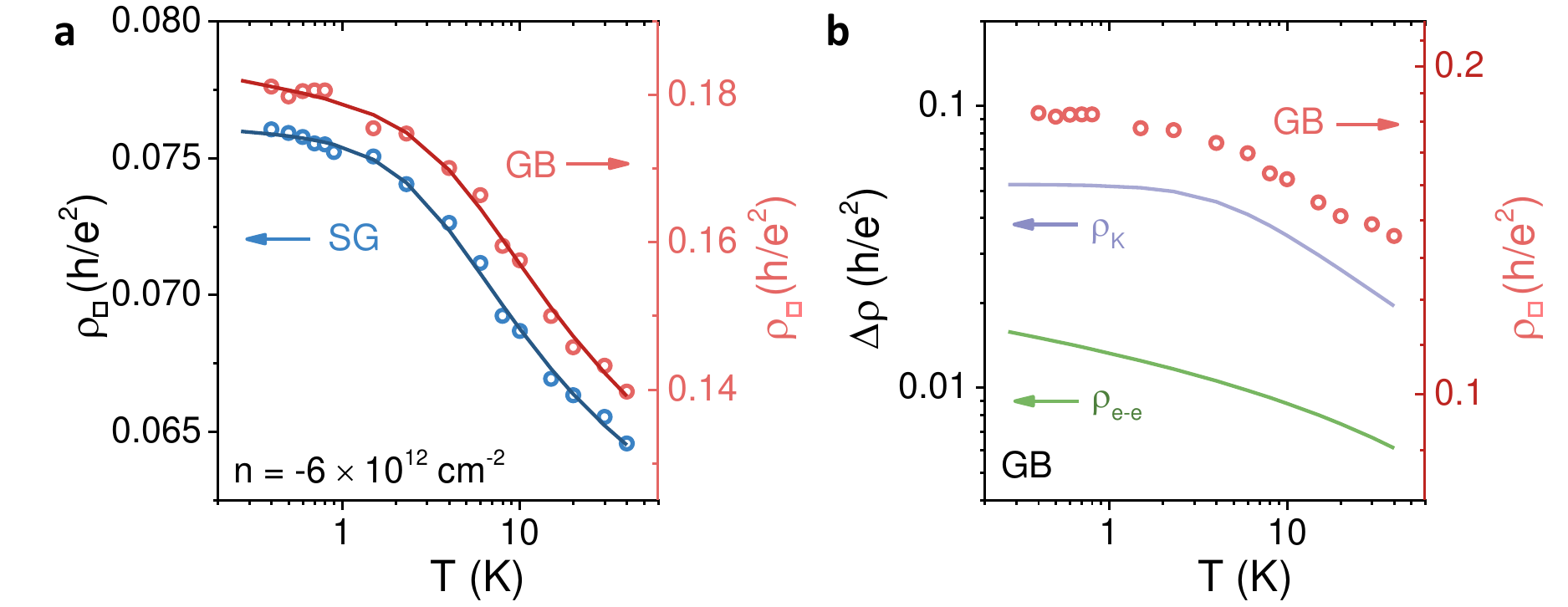}
\caption{\label{rhovsT} \textbf{a},~$T$-dependence of the reduced sheet resistance $\rho_{\square}$ (in units of $h/e^2$) measured at $B=0.5$~T for the SG region (blue circles) and the GB region (red circles) at $n=-6\times10^{12}$~cm$^{-2}$. The solid lines show fit to the Eq.~\ref{rhovsT_eq}, which includes electron-electron interaction corrections $\rho_{e-e}$ and the Kondo contribution to resistivity $\rho_{\mathrm{K}}$. \textbf{b},~The extracted values of $\rho_{e-e}$ (green solid line), $\rho_{\mathrm{K}}$ (purple solid line) and the experimentally obtained $\rho_{\square}$ for the GB region at $n=-6\times10^{12}$~cm$^{-2}$ are plotted as a function of $T$ to show their relative magnitudes.}
%\end{center}
\end{figure*}

%\begin{figure*}[h!]
%\begin{center}
%\includegraphics[width=1\linewidth]{rhovsT_v2.pdf}
%\caption{\label{rhovsT_v2} $T$-dependence of the reduced sheet resistance $\rho_{\square}$ (in units of $h/e^2$) measured at $B=0.5$~T for the SG region at \textbf{a}, $n=1\times10^{12}$~cm$^{-2}$, \textbf{b}, $n=-6\times10^{12}$~cm$^{-2}$, and for the GB region at \textbf{c}, $n=1\times10^{12}$~cm$^{-2}$, \textbf{d}, $n=-6\times10^{12}$~cm$^{-2}$. The solid lines show fit to the Eq.~\ref{rhovsT_eq} accounting for the electron-electron interaction corrections $\rho_{e-e}$ and the Kondo contribution to resistivity $\rho_{\mathrm{K}}$.}
%\end{center}
%\end{figure*}

%\begin{figure*}[h!]
%\begin{center}
%\includegraphics[width=1\linewidth]{rhovsT.pdf}
%\caption{\label{rhovsT} $T$-dependence of the reduced sheet resistance $\rho_{\square}$ (in units of $h/e^2$) including electron-electron interaction corrections $\rho_{e-e}$ and the Kondo contribution to resistivity $\rho_{\mathrm{K}}$, plotted for the GB region at $B=0.5$~T and $n=-6\times10^{12}$~cm$^{-2}$.}
%\end{center}
%\end{figure*}

The Kondo temperature $T_{\mathrm{K}}$ can be extracted from the temperature dependence of the reduced sheet resistance $\rho_{\square}$ ($R_{\square}$ in units of $h/e^2$) at $B=0.5$~T where weak localization effects have been mostly eliminated. Assuming contributions only from a $T$-independent longitudinal resistivity $\rho_{c}$, electron-electron scattering-induced correction $\rho_{e-e}$~\cite{Lara_PRL2011,Jobst_PRL2012} and a Kondo contribution $\rho_{\mathrm{K}}$~\cite{Chen_NatPhys2011,Gordon_PRL1998}:

\begin{eqnarray}\label{rhovsT_eq}
\rho_{\square} &=& \rho_c+\rho_{e-e}+\rho_{\mathrm{K}} \nonumber \\
&=& \rho_c+(\mu^2 B^2 -1) \frac{\rho_{c}^2}{\pi} A \ln\frac{\hbar}{k_{\mathrm{B}} T \tau_{tr}}+\rho_{\mathrm{K,0}} \left[ 1+\left(\frac{T}{T_{\mathrm{K}}}\right)^2 (2^{1/0.22}-1)\right]^{-0.22}
\end{eqnarray}

\noindent where $\mu$ is the device mobility, the coefficient $A=1+c\left[1-\frac{\ln(1+F^{\sigma}_0)}{F^{\sigma}_0}\right]$ is a measure of the interaction strength (where $F^{\sigma}_0$ is the Fermi-liquid constant and $c$ is the number of contributing multiplet channels), $\tau_{tr}$ is the transport time derived from the Drude resistivity, and $\rho_{\mathrm{K}}$ is the Kondo resistivity at zero temperature. Fig.~\ref{rhovsT}a show the fits to Eq.~\ref{rhovsT_eq} for the resistivity versus temperature data of the SG (blue circles) and GB (red circles) regions at $n=-6\times10^{12}$~cm$^{-2}$, keeping $\rho_c$, $\rho_{\mathrm{K,0}}$, $A$ and $T_{\mathrm{K}}$ as fitting parameters. It is evident from Fig.~\ref{rhovsT}a that $\rho_{\square}$ saturates at a slightly higher temperature for the GB region as compared to the SG region, which manifests in a relatively high Kondo temperature $T_{\mathrm{K}}\sim 20$~K). Fitting the resistivity data for the GB region at $n=-6\times10^{12}$~cm$^{-2}$ with Eq.~\ref{rhovsT_eq}, we get $\rho_c\approx 0.11(h/e^2)$, $\rho_{\mathrm{K,0}}\approx 0.05(h/e^2)$, $A\approx 0.47$ and $T_{\mathrm{K}}\approx 20$~K respectively. Finally, Fig.~\ref{rhovsT}b compares the relative contribution of the {\it e-e} interaction-induced corrections $\rho_{e-e}$ (green solid line) and the Kondo contribution $\rho_{\mathrm{K}}$ (purple solid line) to the total experimental value (red circles) obtained for the GB region at $n=-6\times10^{12}$~cm$^{-2}$.

\newpage
\begin{lyxlist}{00.00.0000}
\item [{\textbf{S12. Temperature dependence of noise \boldmath{$N=S_\sigma/\sigma^2$}}}]
\end{lyxlist}

\begin{figure*}[h!]
%\begin{center}
\includegraphics[width=1\linewidth]{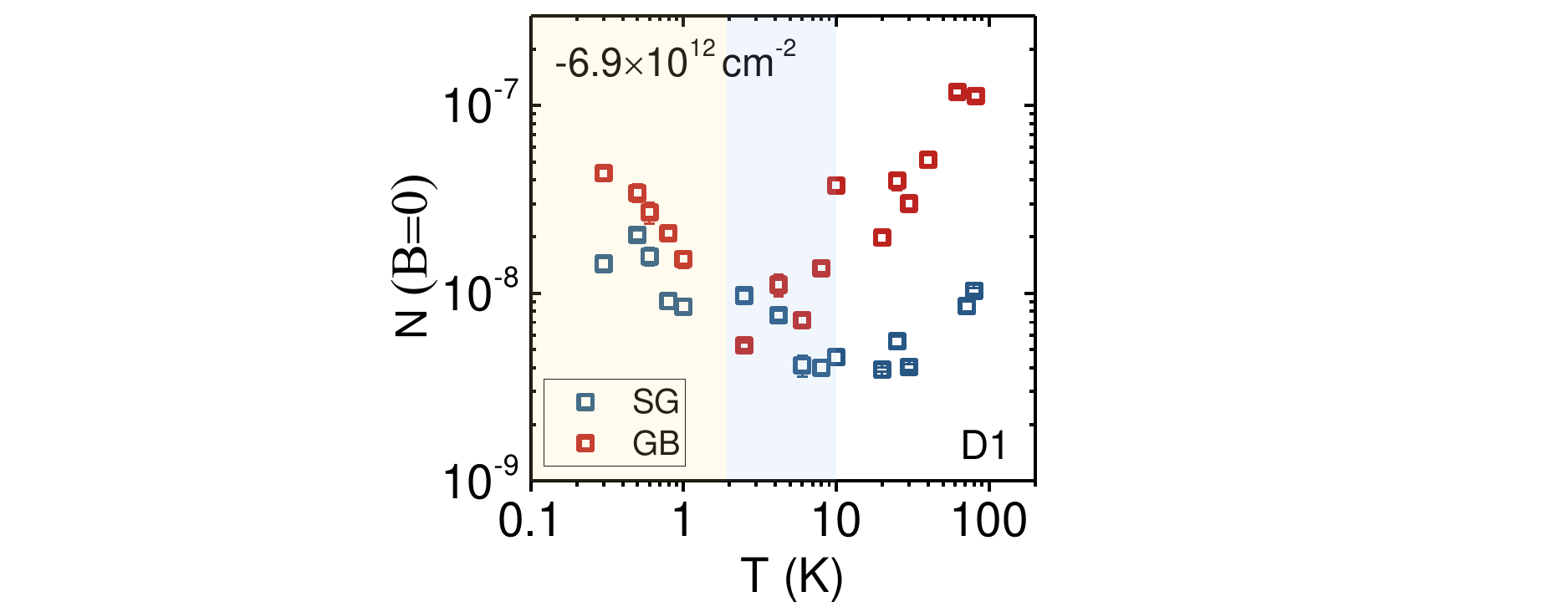}
\caption{\label{NoisevsT} Normalized variance $N=S_\sigma/\sigma^2$ at $B=0$ as a function of temperature is plotted for SG (blue) and GB (red) regions at $n=-6.9\times10^{12}$~cm$^{-2}$ for device D1.}
%\end{center}
\end{figure*}

The noise $N=S_\sigma/\sigma^2$ at $B=0$ in both the SG and GB regions of D1 at $n=-6.9\times10^{12}$~cm$^{-2}$ (Fig.~\ref{NoisevsT}) show clear non-monotonic temperature dependences. The noise initially decreases as the temperature is lowered in an almost linear fashion~($\propto{T}$) as expected from the noise models based on carrier tunneling between the channel and trap states spread over an energy range $\sim k_\mathrm{B} T$ about the Fermi energy. Below $T\approx 10$~K (which matches with the temperature scale in Fig.~S7a), quantum interference contributions can no longer be neglected, and the increase in SG noise with decreasing $T$ can be explained using the Feng-Lee-Stone~(FLS) theory of UCF noise~\cite{Feng_PRL1986}. However, the comparatively sharper increase in noise of the GB region below $T\approx 2$~K cannot be attributed to conventional FLS theory alone. This temperature scale matches with both the observed spontaneous breaking of TRS in magneto-noise and the resistivity downturn in the time-averaged equilibrium resistance, and can be explained through the drastic increase in chaotic spin ensemble configurations in the spin glass phase~\cite{Feng_PRB1987}.

\cleardoublepage{}

\cleardoublepage{}